\documentclass[letterpaper]{CECS}
\usepackage{graphicx}
\usepackage{amsfonts}

\title{(Super)-Gravities Beyond 4 Dimensions\thanks{%
Lectures given at the 2001 Summer School {\it Geometric and
Topological Methods for Quantum Field Theory}, Villa de Leyva,
Colombia, June 2001. E-mail: {\tt  jz@cecs.cl}}}
\author{ Jorge Zanelli \\ Centro de Estudios Cient\'{\i}ficos (CECS), Casilla 1469, Valdivia, Chile.}

\preprint{{\tiny CECS-PHY-02/06} }

\abstract{These lectures are intended as a broad introduction to
Chern Simons gravity and supergravity. The motivation for these
theories lies in the desire to have a gauge invariant action --in
the sense of fiber bundles-- in more than three dimensions, which
could provide a firm ground for constructing a quantum theory of
the gravitational field. The case of Chern-Simons gravity and its
supersymmetric extension for all odd $D$ is presented. No
analogous construction is available in even dimensions.}

\begin{document}
\newpage

\begin{center}
{\bf LECTURE 1} \vskip 0.2cm {\bf GENERAL RELATIVITY REVISITED}
\end{center}
\vskip 0.3cm

In this lecture, the standard construction of the action principle for general
relativity is discussed. The scope of the analysis is to set the basis for a
theory of gravity in any number of dimensions, exploiting the similarity
between gravity and a gauge theory as a fiber bundle. It is argued that in a
theory that describes the spacetime geometry, the metric and affine properties
of the geometry should be represented by independent entities, an idea that
goes back to the works of Cartan and Palatini. I it shown that he need for an
independent description of the affine and metric features of the geometry leads
naturally to a formulation of gravity in terms of two independent 1-form
fields: the vielbein, $e^{a}$, and the spin connection $\omega_{\;b}^{a}$.

Since these lectures are intended for a mixed audience/readership of
mathematics and physics students, it would seem appropriate to locate the
problems addressed here in the broader map of physics.

\section{Physics and Mathematics.}

Physics is an experimental science. Current research, however, especially in
string theory, could be taken as an indication that the experimental basis of
physics is unnecessary. String theory not only makes heavy use of sophisticated
modern mathematics, it has also stimulated research in some fields of
mathematics. At the same time, the lack of direct experimental evidence, either
at present or in the foreseeable future, might prompt the idea that physics
could exist without an experimental basis. The identification, of high energy
physics as a branch of mathematics, however, is only superficial. High energy
physics in general and string theory in particular, have as their ultimate goal
the description of nature, while Mathematics is free from this constraint.

There is, however, a mysterious connection between physics and mathematics
which runs deep, as was first noticed probably by Pythagoras when he concluded
that that, at its deepest level, reality is mathematical in nature. Such is the
case with the musical notes produced by a violin string or by the string that
presumably describes nature at the Planck scale.

Why is nature at the most fundamental level described by simple, regular,
beautiful, mathematical structures? The question is not so much how structures
like knot invariants, the index theorem or moduli spaces appear in string
theory as gears of the machinery, but why should they occur at all. As E.
Wigner put it, ``{\em the miracle of the appropriateness of the language of
mathematics for the formulation of the laws of physics is a wonderful gift,
which we neither understand nor deserve.}''\cite{Wigner}.

Often the connection between theoretical physics and the real world is
established through the phenomena described by solutions of differential
equations. The aim of the theoretical physicist is to provide economic
frameworks to explain why those equations are necessary. The time-honored
approach to obtain dynamical equations is a variational principle: the
principle of least action in Lagrangian mechanics, the principle of least time
in optics, the principle of highest profit in economics, etc. These principles
are postulated with no further justification beyond their success in providing
differential equations that reproduce the observed behavior. However, there is
also an important aesthetic aspect, that has to do with economy of assumptions,
the possibility of a wide range of predictions, simplicity, beauty.

In order to find the correct variational principle, an important criterion is
symmetry. Symmetries are manifest in the conservation laws observed in the
phenomena. Under some suitable assumptions, symmetries are often strong enough
to select the general form of the possible action functionals.

The situation can be summarized more or less in the following scheme:

\begin{center}
{\small
\begin{tabular}{ccccc}
\begin{tabular}{l}
{\bf \tiny Theoretical} \\
{\bf \tiny predictions} \\
$\;\;\;\;\;\;\;\;\;\;\downarrow$
\end{tabular} &
\begin{tabular}{|c|c|c|}
\hline
 {\bf Feature}&{\bf Ingredient} & {\bf Examples}  \\ \hline
 Symmetries & Symmetry& Translations,\\
 &  group &  Lorentz,
gauge \\ \hline
 Variational & Action & $\delta I=\delta \!\!
\int (T \! - \! V)dt$\\
 Principle & Functional & $=0.\;\;\;\;\;\;\;\;\;\;\;\;\;$ \\ \hline
 Dynamics & Field & $ \vec{F}=m
\vec{a}$, \\
 &  Equations &  Maxwell eqs.   \\ \hline Phenomena &
Solutions & Orbits, states  \\
 & & trajectories \\ \hline Experiments &
Data & Positions, times \\ \hline
\end{tabular}
&
\begin{tabular}{l}
$\;\;\;\;\;\;\;\;\;\; \uparrow$ \\
{\bf \tiny Theoretical} \\
{\bf \tiny construction}
\end{tabular}
\end{tabular}
}
\end{center}
\vskip 0.3cm Theoretical research proceeds inductively, upwards from the
bottom, guessing the theory from the experimental evidence. Once a theory is
built, it predicts new phenomena that should be confronted with experiments,
checking the foundations, as well as the consistency of the building above.
Axiomatic presentations, on the other hand, go from top to bottom. They are
elegant and powerful, but they rarely give a clue about how the theory was
constructed and they hide the fact that a theory is usually based on very
little experimental evidence, although a robust theory will generate enough
predictions and resist many experimental tests.

\subsection{Renormalizability and the Success of Gauge Theory}

A good example of this way of constructing a physical theory is provided by
Quantum Field Theory. Experiments in cloud chambers during the first half of
the twentieth century showed collisions and decays of particles whose mass,
charge, and a few other attributes could be determined. From this data, a
general pattern of possible and forbidden reactions as well as relative
probabilities of different processes was painfully constructed. Conservation
laws, selection rules, new quantum numbers were suggested and a
phenomenological model slowly emerged, which reproduced most of the
observations in a satisfactory way. A deeper understanding, however, was
lacking. There was no theory from which the laws could be deduced simply and
coherently. The next step, then, was to construct such a theory. This was a
major enterprise which finally gave us the Standard Model. The humble word
``model'', used instead of ``theory'', underlines the fact that important
pieces are still missing in it.

The model requires a classical field theory described by a lagrangian capable
of reproducing the type of interactions (vertices) and conservation laws
observed in the experiments at the lowest order (low energy, weakly interacting
regime). Then, the final test of the theory comes from the proof of its
internal consistency as a quantum system: {\bf renormalizability}.

It seems that Hans Bethe was the first to observe that non renormalizable
theories would have no predictive power and hence renormalizability should be
the key test for the physical consistency of a theory \cite{Schweber}. A
brilliant example of this principle at work is offered by the theory for
electroweak interactions. As Weinberg remarked in his Nobel lecture, if he had
not been guided by the principle of renormalizability, his model would have
included contributions not only from $SU(2)\times U(1)$-invariant vector boson
interactions --which were believed to be renormalizable, although not proven
until a few years later by 't Hooft \cite{'t Hooft}-- but also from the
$SU(2)\times U(1)$-invariant four fermion couplings, which were known to be non
renormalizable \cite{WeinbergNobel}. Since a non renormalizable theory has no
predictive power, even if it could not be said to be incorrect, it would be
scientifically irrelevant like, for instance, a model based on angels and evil
forces.

One of the best examples of a successful application of mathematics for the
description of nature at a fundamental scale is the principle of {\bf gauge
invariance}, that is the invariance of a system under a symmetry group that
acts locally. The underlying mathematical structure of the gauge principle is
mathematically captured through the concept of {\bf fiber bundle}, as discussed
in the review by Sylvie Paycha in this school \cite{Paycha}. For a discussion
of the physical applications, see also \cite{Nakahara}.

Three of the four forces of nature (electromagnetism, weak, and strong
interactions) are explained and accurately modelled by a Yang-Mills action
built on the assumption that nature should be invariant under a group of
transformations acting independently at each point of spacetime. This local
symmetry is the key ingredient in the construction of physically testable
(renormalizable) theories. Thus, symmetry principles are not only useful in
constructing the right (classical) action functionals, but they are often
sufficient to ensure the viability of a quantum theory built from a given
classical action.

\subsection{The Gravity Puzzle}

The fourth interaction of nature, the gravitational attraction, has stubbornly
resisted quantization. This is particularly irritating as gravity is built on
the principle of invariance under general coordinate transformations, which is
a local symmetry analogous to the gauge invariance of the other three forces.
These lectures will attempt to shed some light on this puzzle.

One could question the logical necessity for the existence of a quantum theory
of gravity at all. True fundamental field theories must be renormalizable;
effective theories need not be, as they are not necessarily described by
quantum mechanics at all. Take for example the Van der Waals force, which is a
residual low energy interaction resulting from the electromagnetic interactions
between electrons and nuclei. At a fundamental level it is all quantum
electrodynamics, and there is no point in trying to write down a quantum field
theory to describe the Van der Waals interaction, which might even be
inexistent. Similarly, gravity could be an effective interaction analogous to
the Van der Waals force, the low energy limit of some fundamental theory like
string theory. There is one difference, however. There is no action principle
to describe the Van der Waals interaction and there is no reason to look for a
quantum theory for molecular interactions. Thus, a biochemical system is not
governed by an action principle and is not expected to be described by a
quantum theory, although its basic constituents are described by QED, which is
a renormalizable theory.

Gravitation, on the other hand, is described by an action principle. This is an
indication that it could be viewed as a fundamental system and not merely an
effective force, which in turn would mean that there might exist a quantum
version of gravity. Nevertheless, countless attempts by legions of researchers
--including some of the best brains in the profession-- through the better part
of the twentieth century, have failed to produce a sensible (e.g.,
renormalizable) quantum theory for gravity.

With the development of string theory over the past twenty years, the
prevailing view now is that gravity, together with the other three interactions
and all elementary particles, are contained as modes of the fundamental string.
In this scenario, all four forces of nature including gravity, would be low
energy effective phenomena and not fundamental reality. Then, the issue of
renormalizabilty of gravity would not arise, as it doesn't in the case of the
Van der Waals force.

Still a puzzle remains here. If the ultimate reality of nature is string theory
and the observed high energy physics is just low energy phenomenology described
by effective theories, there is no reason to expect that electromagnetic, weak
and strong interactions should be governed by renormalizable theories at all.
In fact, one would expect that those interactions should lead to non
renormalizable theories as well, like gravity or the old four-fermion model for
weak interactions. If these are effective theories like thermodynamics or
hydrodynamics, one could even wonder why these interactions are described by an
action principle at all.

\subsection{Minimal Couplings and Connections}

Gauge symmetry fixes the form in which matter fields couple to the carriers of
gauge interactions. In electrodynamics, for example, the ordinary derivative in
the kinetic term for the matter fields, $\partial_{\mu }$, is replaced by the
covariant derivative,
\begin{equation}
\nabla _{\mu }=\partial _{\mu }+A_{\mu }. \label{minimal}
\end{equation}
This provides a unique way to couple charged fields, like the electron, and the
electromagnetic field. At the same time, this form of interaction avoids
dimensionful coupling constants in the action. In the absence of such coupling
constants, the perturbative expansion is likely to be well behaved because
gauge symmetry imposes severe restrictions on the type of terms that can be
added to the action, as there are very few gauge invariant expressions in a
given number of spacetime dimensions. Thus, if the Lagrangian contains all
possible terms allowed by the symmetry, perturbative corrections could only
lead to rescalings of the coefficients in front of each term in the Lagrangian.
These rescalings can always be absorbed in a redefinition of the parameters of
the action. This {\bf renormalization} procedure that works in gauge theories
is the key to their internal consistency.

The ``vector potential'' $A_{\mu }$ is a connection 1-form, which means that,
under a gauge transformation,
\begin{equation}
{\bf A(x)} \rightarrow {\bf A(x)}^{\prime}= {\bf U}(x){\bf
A(x)U}(x)^{-1} + {\bf U}(x)d{\bf U}^{-1}(x), \label{gauge}
\end{equation}
where $U(x)$ represents a position dependent group element. The value of ${\bf
A}$ depends on the choice of gauge ${\bf U}(x)$ and it can even be made to
vanish at a given point by an appropriate choice of $U(x)$. The combination
$\nabla_{\mu }$ is the covariant derivative, a differential operator that,
unlike the ordinary derivative and ${\bf A}$ itself, transforms homogeneously
under the action of the gauge group,
\begin{equation}
\nabla_{\mu} \rightarrow \nabla_{\mu }^{\prime}={\bf U}(x)\nabla_{\mu}.
\end{equation}

The connection can in general be a matrix-valued object, as in the case of
nonabelian gauge theories. In that case, ${\bf \nabla }_{\mu}$ is an operator
1-form,
\begin{eqnarray}
{\bf \nabla} &=&d+{\bf A}  \label{covariantDer-1} \\
&=&dx^{\mu }(\partial _{\mu }+{\bf A}_{\mu }).  \nonumber
\end{eqnarray}
Acting on a function $\phi(x)$, which is in a vector representation of the
gauge group ($\phi(x)\rightarrow \phi^{\prime }(x)={\bf U}(x)\cdot \phi(x)$),
the covariant derivative reads
\begin{equation}
{\bf \nabla}\phi = d\phi + {\bf A}\wedge\phi.\label{delphi}
\end{equation}
The covariant derivative operator ${\bf \nabla}$ has a remarkable property: its
square is not a differential operator but a multiplicative one, as can be seen
from (\ref{delphi})
\begin{eqnarray}
{\bf \nabla}{\bf \nabla}\phi&=&d({\bf A}\phi)+ {\bf A}d\phi+{\bf A\wedge A}\phi
\\ \nonumber &=& (d {\bf A}+{\bf A}\wedge{\bf A})\phi \\ \nonumber &=& {\bf
F}\phi
\end{eqnarray}
The combination ${\bf F}= d{\bf A} + {\bf A}\wedge{\bf A}$ is the field
strength of the nonabelian interaction. This generalizes the electric and
magnetic fields of electromagnetism and it indicates the presence of energy.

One can see now why the gauge principle is such a powerful idea in physics: the
covariant derivative of a field, ${\bf \nabla}\phi$, defines the coupling
between $\phi$ and the gauge potential ${\bf A}$ in a unique way. Furthermore,
${\bf A}$ has a uniquely defined field strength ${\bf F}$, which in turn
defines the dynamical properties of the gauge field. In 1954, Robert Mills and
Chen-Nin Yang grasped the beauty and the power of this idea and constructed
what has been since known as the nonabelian Yang-Mills theory
\cite{Yang-Mills}.

On the tangent bundle, the covariant derivative corresponding to the gauge
group of general coordinate transformations is the usual covariant derivative
in differential geometry,
\begin{eqnarray}
{\bf D} &=&d+{\bf \Gamma }  \label{covariantDer-2}
\\ &=&dx^{\mu }(\partial _{\mu }+{\bf \Gamma}_{\mu }),  \nonumber
\end{eqnarray}
where ${\bf \Gamma }$ is the Christoffel symbol, involving the metric and its
derivatives.

The covariant derivative operator in both cases reflects the fact that these
theories are invariant under a group of local transformations, that is,
operations which act independently at each point in space. In electrodynamics
${\bf U}(x)$ is an element of $U(1)$, and in the case of gravity ${\bf U}(x)$
is the Jacobian matrix $(\partial x/\partial x^{\prime})$, which describes a
diffeomorphism, or general coordinate change, $x\rightarrow x^{\prime }$.

\subsection{Gauge Symmetry and Diffeomorphism Invariance}

The close analogy between the covariant derivatives ${\bf \nabla}$ and ${\bf
D}$ could induce one to believe that the difficulties for constructing a
quantum theory for gravity shouldn't be significantly worse than for an
ordinary gauge theory like QED. It would seem as if the only obstacles one
should expect would be technical, due to the differences in the symmetry group,
for instance. There is, however, a more profound difference between gravity and
the standard gauge theories that describe Yang-Mills systems. The problem is
not that General Relativity lacks the ingredients to make a gauge theory, but
that the right action for gravity in four dimensions cannot be written as that
of a gauge invariant system for the diffeomorphism group.

In a YM theory, the connection ${\bf A}_{\mu }$ is an element of a Lie algebra
whose structure is independent of the dynamical equations. In electroweak and
strong interactions, the connection is a dynamical field, while both the base
manifold and the symmetry group are fixed, regardless of the values of the
connection or the position in spacetime. This implies that the Lie algebra has
structure constants, which are neither functions of the field $A$, or the
position $x$. If $G^{a}(x)$ are the gauge generators in a YM theory, they obey
an algebra of the form
\begin{equation}
\lbrack G^{a}(x),G^{b}(y)]=C_{c}^{ab}\delta (x,y)G^{c}(x),  \label{LieG}
\end{equation}
where $C_{c}^{ab}$ are the structure constants.

The Christoffel connection ${\bf \Gamma }_{\beta \gamma }^{\alpha }$, instead,
represents the effect of parallel transport over the spacetime manifold, whose
geometry is determined by the dynamical equations of the theory. The
consequence of this is that the diffeomorphisms do not form a Lie algebra but
an {\bf open algebra}, which has {\em structure functions} instead of {\em
structure constants} \cite{Henneaux}. This problem can be seen explicitly in
the diffeomorphism algebra generated by the hamiltonian constraints of gravity,
${\cal H}_{\perp }(x)$, ${\cal H}_{i}(x)$,
\begin{equation}
\begin{array}{lll}
\lbrack {\cal H}_{\perp }(x),{\cal H}_{\perp }(y)] & = & g^{ij}(x)
\delta(x,y),_{i}{\cal H}_{j}(y)-g^{ij}(y)\delta (y,x),_{i}{\cal H}_{j}(x) \\
\lbrack {\cal H}_{i}(x),{\cal H}_{j}(y)] & = & \delta
(x,y),_{i}{\cal H}_{j}(y)-\delta (x,y),_{j}{\cal H}_{i}(y) \\
\lbrack {\cal H}_{\perp }(x),{\cal H}_{i}(y)] & = &
\delta(x,y),_{i}{\cal H}_{\perp }(y)
\end{array}
,  \label{LieH}
\end{equation}
where $\delta (y,x),_{i}=\frac{\partial \delta (y,x)}{\partial x^{i}}$.

Here one now finds functions of the dynamical fields, $g^{ij}(x)$ playing the
role of the structure constants $C_{c}^{ab}$, which identify the symmetry group
in a gauge theory. If the structure ``constants'' were to change from one point
to another, it would mean that the symmetry group is not uniformly defined
throughout spacetime, which would prevent an interpretation of gravity in terms
of fiber bundles, where the base is spacetime and the symmetry group is the
fiber.

It is sometimes asserted in the literature that gravity is a gauge theory for
the translation group, much like the Yang Mills theory of strong interactions
is a gauge theory for the $SU(3)$ group. We see that although this is
superficially correct, the usefulness of this statement is limited by the
profound differences a gauge theory with fiber bundle structure and another
with an open algebra such as gravity.

\section{General Relativity}

The question we would like to address is: {\it What would you say is the right
action for the gravitational field in a spacetime of a given dimension? }On
November 25 1915, Albert Einstein presented to the Prussian Academy of Natural
Sciences the equations for the gravitational field in the form we now know as
Einstein equations \cite{Einstein}. Curiously, five days before, David Hilbert
had proposed the correct action principle for gravity, based on a communication
in which Einstein had outlined the general idea of what should be the form of
the equations \cite{Hilbert}. This is not so surprising in retrospect, because
as we shall see, there is a unique action in four dimensions which is
compatible with general relativity that has flat space as a solution. If one
allows nonflat geometries, there is essentially a one-parameter family of
actions that can be constructed: the Einstein-Hilbert form plus a cosmological
term,
\begin{equation} \label{EinsteinAction}
I[g]=\int \sqrt{-g}(\alpha _{1}R+\alpha _{2})d^{4}x,
\end{equation}
where $R$ is the scalar curvature, which is a function of the metric $g_{\mu
\nu }$, its inverse $g^{\mu \nu }$, and its derivatives (for the definitions
and conventions we use here, see Ref.\cite{MTW} ). The expression $I[g]$ is the
only functional of the metric which is invariant under general coordinate
transformations and gives second order field equations in four dimensions. The
coefficients $\alpha _{1}$ and $\alpha _{1}$ are related to the gravitational
constant and the cosmological constant through
\begin{equation}
\alpha _{1}=\frac{1}{16\pi G}\;,\;\alpha _{2}=\frac{\Lambda }{ 8\pi G}.
\label{G-Constants}
\end{equation}

Einstein equations are obtained by extremizing this action
(\ref{EinsteinAction}) and they are unique in that:

({\bf i}) They are tensorial equations

({\bf ii}) They involve only up to second derivatives of the metric

({\bf iii}) They reproduce Newtonian gravity in the weak field nonrelativistic
approximation.

The first condition implies that the equations have the same meaning in all
coordinate systems. This follows from the need to have a coordinate independent
(covariant) formulation of gravity in which the gravitational force is replaced
by the nonflat geometry of spacetime. The gravitational field being a
geometrical entity implies that it cannot resort to a preferred coordinate
choice or, in physical terms, a preferred set of observers.

The second condition means that Cauchy conditions are necessary (and sufficient
in most cases) to integrate the equations. This condition is a concession to
the classical physics tradition: the possibility of determining the
gravitational field at any moment from the knowledge of the positions and
momenta at a given time. This requirement is also the hallmark of Hamiltonian
dynamics, which is the starting point for canonical quantum mechanics.

The third requirement is the correspondence principle, which accounts for our
daily experience that an apple and the moon fall the the way they do.

If one further assumes that Minkowski space be among the solutions of the
matter-free theory, then one must set $\Lambda =0$, as most sensible particle
physicists would do. If, on the other hand, one believes in static homogeneous
and isotropic cosmologies, then $\Lambda$ must have a finely tuned nonzero
value. Experimentally, $\Lambda$ has a value of the order of $10^{-120}$ in
some ``natural'' units \cite{Weinberg}. Furthermore, astrophysical measurements
seem to indicate that $\Lambda $ must be positive \cite{LambdaExp}. This
presents a problem because there seems to be no theoretical way to predict this
``unnaturally small'' nonzero value.

As we will see in the next lecture, for other dimensions, the Einstein-Hilbert
action is not the only possibility in order to satisfy conditions ({\bf
i-iii}).

\subsection{Metric and Affine Structures}

We conclude this introduction by discussing what we mean by spacetime geometry.
Geometry is sometimes understood as the set of assertions one can make about
the points in a manifold and their relations. This broad (and vague) idea, is
often interpreted as encoded in the metric tensor, $g_{\mu \nu }(x)$, which
provides the notion of distance between nearby points with slightly different
coordinates,
\begin{equation}
ds^{2}=g_{\mu \nu }\;dx^{\mu }dx^{\nu }.  \label{Metric}
\end{equation}

This is the case in Riemannian geometry, where all objects that are relevant
for the spacetime can be constructed from the metric. However, one can
distinguish between {\bf metric} and {\bf affine} features of space, that is,
between the notions of {\bf distance} and{\bf \ parallelism}. Metricity refers
to lengths, areas, volumes, etc., while affinity refers to scale invariant
properties such as shapes.

Euclidean geometry was constructed using two elementary instruments: the
compass and the (unmarked) straightedge. The first is a metric instrument
because it allows comparing lengths and, in particular, drawing circles. The
second is used to draw straight lines which, as will be seen below, is a basic
affine operation. In order to fix ideas, let's consider a few examples from
Euclidean geometry. Pythagoras' famous theorem is a metric statement; it
relates the lengths of the sides of a triangle:

\begin{figure}[ht]
\begin{center}
\includegraphics[height=4cm]{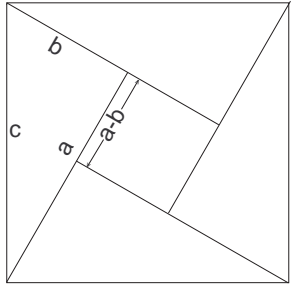}
\caption{Pythagoras theorem: ${\sf c}^2={\sf (a-b)}^2+4 {\sf [a
b]}/2$}
\end{center}
\end{figure}

Affine properties on the other hand, do not change if the length scale is
changed, such as the shape of a triangle or, more generally, the angle between
two straight lines. A typical affine statement is, for instance, the fact that
when two parallel lines intersect a third, the corresponding angles are equal,
as seen in {\bf Fig.2}.

\begin{figure}[ht]
\begin{center}
\includegraphics[height=4cm]{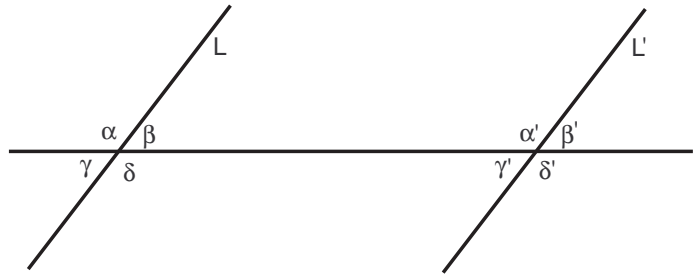}
\caption{Affine property: {\sf L} $\parallel$ {\sf L}$^{\prime }
\Leftrightarrow \alpha ={\alpha}^{\prime}=\delta =
{\delta}^{\prime}
 \;,\; \beta ={\beta }^{\prime } =\gamma ={\gamma }^{\prime } $}
\end{center}
\end{figure}
\smallskip
Of course parallelism can be reduced to metricity. As we learned in school, one
can draw a parallel to a line {\bf L} using a right angled triangle ({\bf W})
and an unmarked straightedge ({\bf R}): One aligns one of the short sides of
the triangle with the straight line and rests the other short side on the
ruler. Then, one slides the triangle to where the parallel is to be drawn, as
in {\bf Fig.3}.

\begin{figure}[ht]
\begin{center}
\includegraphics[height=6cm]{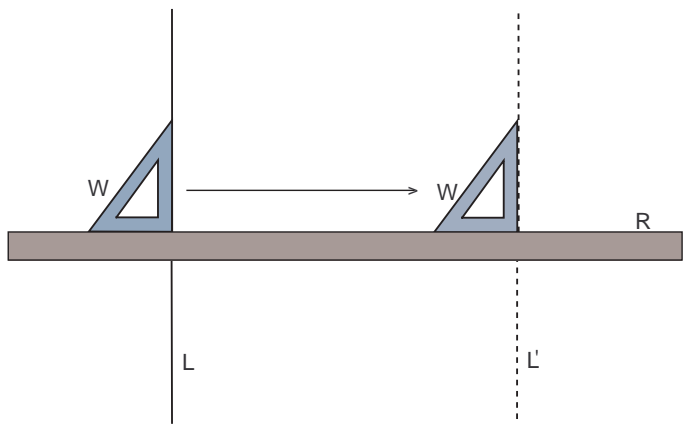}
\caption{ Constructing parallels using a right-angled triangle ({\sf W}) and a
straightedge ({\sf R}) }
\end{center}
\end{figure}

Thus, given a way to draw right angles and a straight line in space, one can
define parallel transport. As any child knows from the experience of stretching
a string or a piece of rubber band, a straight line is the shape of the
shortest line between two points. This is clearly a metric feature because it
requires {\em measuring} lengths. Orthogonality is also a metric notion that
can be defined using the scalar product obtained from the metric. A right angle
is a metric feature because we should be able to {\em measure} angles, or
measure the sides of triangles\footnote{The Egyptians knew how to {\em use}
Pythagoras' theorem to make a right angle, although they didn't know how to
prove it. Their recipe was probably known long before, and all good
construction workers today still know the recipe: make a loop of rope with 12
segments of equal length. Then, the triangle formed with the loop so that its
sides are 3, 4 and 5 segments long is such that the shorter segments are
perpendicular to each other \cite{Dunham}.}. We will now show that, strictly
speaking, {\em parallelism does not require metricity}.

There is something excessive about the construction in {\bf Fig.3} because one
doesn't {\em have} to use a right angle. In fact, any angle could be used in
order to draw a parallel to {\bf L} in the last example, so long as it {\em
doesn't change} when we slide it from one point to another, as shown in {\bf
Fig.4}.

\begin{figure}[ht]
\begin{center}
\includegraphics[height=5cm]{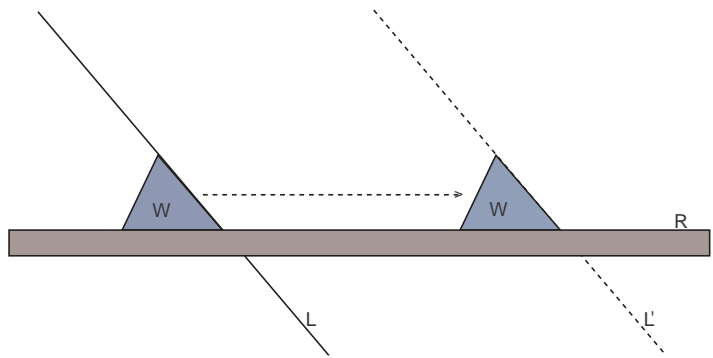}
\caption{ Constructing parallels using an arbitrary angle-preserving  wedge
({\sf W}) and a straightedge ({\sf R}) .}
\end{center}
\end{figure}

We see that the essence of parallel transport is a rigid, angle-preserving
wedge and a straightedge to connect two points. There is still some cheating in
this argument because we took the construction of a straightedge for granted.
What if we had no notion of distance, how do we know what a straight line is?

There is a way to construct a straight line that doesn't require a notion of
distance between two points in space. Take two short enough segments (two short
sticks, matches or pencils would do), and slide them one along the other, as a
cross country skier would do. In this way a straight line is generated by {\em
parallel transport of a vector along itself}, and we have not used distance
anywhere. It is this {\em affine} definition of a straight line that can be
found in Book I of Euclid's Elements. This definition could be regarded as the
{\em straightest line}, which does not necessarily coincide with the{\em \ line
of shortest distance}. They are conceptually independent.

In a space devoid of a metric structure the straightest line could be a rather
strange looking curve, but it could still be used to define parallelism.
Suppose the ruler {\bf R} has been constructed by transporting a vector along
itself, then one can use it to define parallel transport as in {\bf Fig.5}.

\begin{figure}[ht]
\begin{center}
\includegraphics[height=6cm]{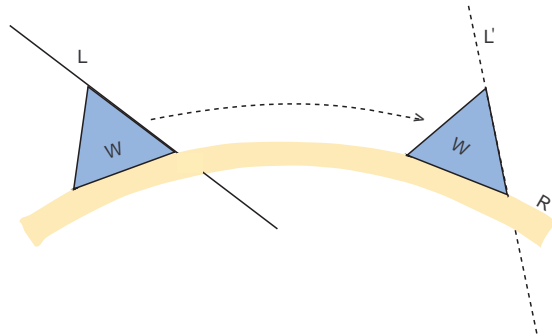}
\caption{ Constructing parallels using any angle-preserving wedge ({\sf W}) and
an arbitrary ruler ({\sf R}). Any ruler is as good as another.}
\end{center}
\end{figure}

There is nothing wrong with this construction apart from the fact that it need
not coincide with the more standard metric construction in {\bf Fig.3}. The
fact that this purely affine construction is logically acceptable means that
parallel transport needs not be a metric concept unless one insists on reducing
affinity to metricity.

In differential geometry, parallelism is encoded in the affine connection
mentioned earlier,$\Gamma _{\beta \gamma }^{\alpha }(x)$, so that a vector $u$
at the point of coordinates $x$ is said to be parallel to the vector
$\tilde{u}$ at a point with coordinates $x+dx$, if their components are related
by ``parallel transport'',
\begin{equation}
\tilde{u}^{\alpha }(x+dx)= \Gamma_{\beta \gamma }^{\alpha
}dx^{\beta }u^{\gamma }(x). \label{Affine}
\end{equation}

The affine connection $\Gamma _{\beta \gamma }^{\alpha }(x)$ need not be
logically related to the metric tensor $g_{\mu \nu }(x)$.

Einstein's formulation of General Relativity adopted the point of view that the
spacetime metric is the only dynamically independent field, while the affine
connection is a function of the metric given by the Christoffel symbol,
\begin{equation}
\Gamma _{\beta \gamma }^{\alpha }=\frac{1}{2}g^{\alpha \lambda }(\partial
_{\beta }g_{\lambda \gamma }+\partial _{\gamma }g_{\lambda \beta }+\partial
_{\lambda }g_{\beta \gamma }).  \label{Christoffel}
\end{equation}

This is the starting point for a controversy between Einstein and Cartan, which
is vividly recorded in the correspondence they exchanged between May 1929 and
May 1932 \cite{Cartan-Einstein}. In his letters, Cartan insisted politely but
forcefully that metricity and parallelism could be considered as independent,
while Einstein pragmatically replied that since the space we live in seems to
have a metric, it would be more economical to assume the affine connection to
be a function of the metric. Einstein argued in favor of economy of independent
fields. Cartan advocated economy of assumptions.

Here we adopt Cartan's point of view. It is less economical in dynamical
variables but is more economical in assumptions and therefore more general.
This alone would not be sufficient argument to adopt Cartan's philosophy, but
it turns out to be more transparent in many ways and to lend itself better to
make a gauge theory of gravity.

\section{First Order Formulation for Gravity}

We view spacetime as a smooth $D$-dimensional manifold of lorentzian signature
$M$, which at every point $x$ possesses a $D$-dimensional tangent space
$T_{x}$. The idea is that this tangent space $T_{x}$ is a good linear
approximation of the manifold $M$ in the neighborhood of $x$. This means that
there is a way to represent tensors over $M$ by tensors on the tangent
space\footnote{Here, only the essential ingredients are given. For a more
extended discussion, there are several texts such as those of
Refs.\cite{Nakahara}, \cite{Schutz} and \cite{Goeckeler-Schuecker} .}.

\subsection{The Vielbein}

The precise translation (isomorphism) between the tensor spaces on $M$ and on
$T_{x}$ is made by means of a dictionary, also called {\it ``soldering form''}
or simply, {\it ``vielbein''}. The coordinate separation $dx^{\mu }$, between
two infinitesimally close points on $M$ is mapped to the corresponding
separation $dz^{a}$ in $T_{x}$, as
\begin{equation}
dz^{a}=e_{\mu }^{a}(x)dx^{\mu }  \label{vielbein}
\end{equation}

The family $\{ e_{\mu }^{a}(x), a=1,...,D=dim M \}$ can also be seen as a local
orthonormal frame on $M$. The definition (\ref{vielbein}) makes sense only if
the vielbein $e_{\mu }^{a}(x)$ transforms as a covariant vector under
diffeomorphisms on $M $ and as a contravariant vector under local Lorentz
rotations of $T_{x}$, $SO(1,D-1)$ (we assumed the signature of the manifold $M$
to be Lorentzian). A similar one to one correspondence can be established
between tensors on $M$ and on $T_{x}$: if $\Pi $ is a tensor with components
$\Pi ^{\mu _{1}...\mu _{n}}$ on $M$, then the corresponding tensor on the
tangent space $T_{x}$ is\footnote{The inverse vielbein $e_{a}^{\mu }(x)$ where
$e_{a}^{\nu }(x)e_{\nu }^{b}(x)=\delta _{a}^{b}$, and $e_{a}^{\nu }(x)e_{\mu
}^{a}(x)=\delta _{\mu }^{\nu }$, relates lower index tensors,
\[
P_{a_{1}...a_{n}}(x)=e_{a_{1}}^{\mu_{1}}(x)\cdot \cdot \cdot
e_{a_{n}}^{\mu _{n}}(x)\Pi _{\mu _{1}...\mu _{n}}(x).
\]}
\begin{equation}
P^{a_{1}...a_{n}}(x)=e_{\mu _{1}}^{a_{1}}(x)\cdot \cdot \cdot e_{\mu
_{n}}^{a_{n}}(x)\Pi ^{\mu _{1}...\mu _{n}}(x).  \label{T-map}
\end{equation}

An example of this map between tensors on $M$ and on $T_{x}$ is the relation
between the metrics of both spaces,
\begin{equation}
g_{\mu \nu }(x)=e_{\mu }^{a}(x)e_{\nu }^{b}(x)\eta _{ab}.  \label{metric}
\end{equation}
This relation can be read as to mean that the vielbein is in this sense the
square root of the metric. Given $e_{\mu }^{a}(x)$ one can find the metric and
therefore, all the metric properties of spacetime are contained in the
vielbein. The converse, however, is not true: given the metric, there exist
infinitely many choices of vielbein that reproduce the same metric. If the
vielbein are transformed as
\begin{equation}
e_{\mu }^{a}(x)\longrightarrow e_{\mu}^{\prime
a}(x)=\Lambda_{b}^{a}(x)e_{\mu }^{b}(x),  \label{Transf-e}
\end{equation}
where the matrix ${\bf \Lambda}(x)$ leaves the metric in the tangent space
unchanged,
\begin{equation}
\Lambda_{c}^{a}(x)\Lambda_{d}^{b}(x)\eta_{ab}=\eta_{cd},
\label{Lorentz}
\end{equation}
then the metric $g_{\mu \nu }(x)$ is clearly unchanged. The matrices that
satisfy (\ref{Lorentz}) form the Lorentz group $SO(1,D-1)$. This means, in
particular, that there are many more components in $e_{\mu }^{a}$ than in
$g_{\mu \nu }$. In fact, the vielbein has $D^{2}$ independent components,
whereas the metric has only $D(D+1)/2$. The mismatch is exactly $D(D-1)/2$, the
number of independent rotations in $D$ dimensions.

\subsection{The Lorentz Connection}

The Lorentz group acts on tensors at each $T_{x}$ independently, that is, the
matrices ${\bf \Lambda}$ that describe the Lorentz transformations are
functions of $x$. In order to define a derivative of tensors in $T_{x}$, one
must compensate for the fact that at neighboring points the Lorentz rotations
are not the same. This is not different from what happens in any other gauge
theory: one needs to introduce a connection for the Lorentz group,
$\omega_{\;b\mu }^{a}(x)$, such that, if $\phi^{a}(x)$ is a field that
transforms as a vector under the Lorentz group, its covariant derivative,
\begin{equation}
D_{\mu}\phi^{a}(x)=\partial _{\mu }\phi^{a}(x)+\omega _{\;b\mu}^{a}(x)
\phi^{b}(x),  \label{D-mu}
\end{equation}
also transforms like a vector under $SO(1,D-1)$ at $x$. This requirement means
that under a Lorentz rotation $\Lambda _{c}^{a}(x)$, $\omega _{\;b\mu }^{a}(x)$
changes as a connection [see (\ref{gauge})]
\begin{equation}
\omega _{\;b\mu }^{a}(x)\longrightarrow \omega_{\;b\mu }^{\prime a}(x)=
\Lambda_{c}^{a}(x)\Lambda _{b}^{d}(x)\omega_{\;d\mu }^{c}(x)+\Lambda
_{c}^{a}(x)\partial _{\mu }\Lambda_{b}^{c}(x). \label{LorConn}
\end{equation}

In physics, $\omega _{\;b\mu }^{a}(x)$ is often called the {\em spin
connection}, but Lorentz connection would be a more appropriate name. The word
``spin" is due to the fact that $\omega _{\;b\mu }^{a}$ arises naturally in the
discussion of spinors, which carry a special representation of the group of
rotations in the tangent space.

The spin connection can be used to define {\em parallel transport} of Lorentz
tensors in the tangent space $T_{x}$ as one goes from the point $x$ to a nearby
point $x+dx$. The parallel transport of the vector field $\phi^{a}(x)$ from the
point $x$ to $x+dx$, is a vector at $x+dx$, $\phi_{||}^{a}(x+dx)$, defined as
\begin{equation}
\phi_{||}^{a}(x+dx)\equiv \phi^{a}(x)+dx^{\mu }\partial_{\mu}
\phi^{a}(x)+dx^{\mu }\omega _{\;b\mu }^{a}(x)\phi^{b}(x).
\label{ParallelTransp}
\end{equation}
Here one sees that the covariant derivative measures the change in a tensor
produced by parallel transport between neighboring points,
\begin{equation}
dx^{\mu }D_{\mu }\phi^{a}(x)=\phi_{||}^{a}(x+dx)-\phi^{a}(x).
\end{equation}
In this way, the affine properties of space are encoded in the components
$\omega_{\;b\mu}^{a}(x)$, which are, until further notice, totally arbitrary
and independent from the metric.

The number of independent components of $\omega _{\;b\mu }^{a}$ is determined
by the symmetry properties of $\omega_{\;\,\;\mu}^{ab} =\eta^{bc}\omega_{\;c\mu
}^{a}$ under permutations of $a$ and $b$. It is easy to see that demanding that
the metric $\eta ^{ab}$ remain invariant under parallel transport implies that
the connection should be antisymmetric, $\omega_{\;\,\;\mu }^{ab} =-\omega
_{\;\, \;\mu }^{ba}$. We leave the proof as an exercise to the reader. Then,
the number of independent components of $\omega _{\;\,\;\mu }^{ab}$ is
$D^{2}(D-1)/2$. This is{\em \ less} than the number of independent components
of the Christoffel symbol, $D^{2}(D+1)/2$.

\subsection{Differential forms}

It can be observed that both the vielbein and the spin connection appear
through the combinations
\begin{equation}
e^{a}(x)\equiv e_{\mu }^{a}(x)dx^{\mu },  \label{1Form-e}
\end{equation}
\begin{equation}
\omega _{\;b}^{a}(x)\equiv \omega _{\;b\mu }^{a}(x)dx^{\mu },
\label{1Form-w}
\end{equation}
that is, they are local 1-forms. This is not an accident. It turns out that all
the geometric properties of $M$ can be expressed with these two 1-forms and
their exterior derivatives only. Since both $e^{a}$ and $\omega^{a}_{\;b}$ only
carry Lorentz indices, these 1-forms are scalars under diffeomorphisms on $M$,
indeed, they are coordinate-free, as all exterior forms. This, means that in
this formalism the spacetime tensors are replaced by tangent space tensors. In
particular, the Riemann curvature 2-form is\footnote{ Here $d$ stands for the
1-form exterior derivative operator $dx^{\mu }\partial_{\mu }\wedge $ .}
\begin{eqnarray}
R_{\;b}^{a} &=& d\omega _{\;b}^{a}+\omega _{\;b}^{a}\wedge
\omega_{\;b}^{a} \nonumber \\
&=&\frac{1}{2}R_{\;b\mu \nu }^{a}dx^{\mu }\wedge dx^{\nu },
\label{Curvature}
\end{eqnarray}
where $R_{\;b\mu \nu }^{a}\equiv e_{\;\alpha }^{a}e_{\; b}^{\beta}R_{\;\beta
\mu \nu }^{\alpha}$ are the components of the usual Riemann tensor projected on
the tangent space (see \cite{MTW}).

The fact that $\omega _{\;b}^{a}(x)$ is a 1-form, just like the gauge potential
in Yang-Mills theory, $A_{\;b}^{a}=A_{\;b\mu }^{a}dx^{\mu }$, suggests that
they are similar, and in fact they both are connections of a gauge
group\footnote{ In what for physicists is fancy language, ${\bf \omega}$ is a
locally defined Lie algebra valued 1-form on $M$, which is also a connection on
the principal $SO(D-1,1)$-bundle over $M$.}. Their transformation laws have the
same form and the curvature $R_{\;b}^{a}$ is completely analogous to the field
strength in Yang-Mills,
\begin{equation}
F_{\;b}^{a}=dA_{\;b}^{a}+A_{\;c}^{a}\wedge A_{\; b}^{c}.
\label{FStrength}
\end{equation}
There is an asymmetry with respect to the vielbein, though. Its transformation
law under the Lorentz group is not that of a connection but of a vector. There
is another important geometric object obtained from derivatives of $e^{a}$
which is analogous to the Riemann tensor is another, the Torsion 2-form,
\begin{equation}
T^{a}=de^{a}+\omega _{\;b}^{a}\wedge e^{b},  \label{Torsion}
\end{equation}
which, unlike $R_{\;b}^{a}$ is a covariant derivative of a vector, and is not a
function of the vielbein only.

Thus, the basic building blocks of first order gravity are $e^{a}$,
$\omega_{\;b}^{a}$, $R_{\;b}^{a}$, $T^{a}$. With them one must put together an
action. But, are there other building blocks? The answer is no and the proof is
by exhaustion. As a cowboy would put it, if there were any more of them 'round
here, we would have heard... And we haven't.

There is a more subtle argument to rule out the existence of other building
blocks. We are interested in objects that transform in a controlled way under
Lorentz rotations (vectors, tensors, spinors, etc.). Taking the covariant
derivatives of $e^{a}$, $R_{\;b}^{a}$, and $T^{a}$, one finds always
combinations of the same objects, or zero:
\begin{eqnarray}
De^{a} &=&de^{a}+\omega _{\;b}^{a}\wedge e^{b}=T^{a}  \label{Torsion2}
\\
DR_{\;b}^{a} &=&dR_{\;b}^{a} + \omega_{\;c}^{a}\wedge R_{\;b}^{c}
+ \omega_{\;b}^{c}\wedge R_{\;c}^{a}=0
\label{Bianchi1} \\
DT^{a} &=&dT^{a}+\omega_{\;b}^{a}\wedge T^{b}=R_{\;b}^{a}\wedge
e^{b}.  \label{Bianchi2}
\end{eqnarray}
The first relation is just the definition of torsion and the other two are the
Bianchi identities, which are directly related to the fact that the exterior
derivative is nilpotent, $d^{2}=\partial _{\mu }\partial _{\nu }dx^{\mu }\wedge
dx^{\nu }=0$. We leave it to the reader to prove these identities.

In the next lecture we discuss the construction of the possible actions for
gravity using these ingredients. In particular, in 4 dimensions, the Einstein
action can be written as
\begin{equation}
I[g]=\int \epsilon_{abcd}(\alpha R^{ab}e^{c}e^{d} + \beta
e^{a}e^{b}e^{c}e^{d}). \label{Einstein1st}
\end{equation}
This is basically the only action for gravity in dimension four, but many more
options exist in higher dimensions.

\vskip 0.5 cm
\begin{center}
{\bf LECTURE 2}
\vskip 0.2 cm
{\bf GRAVITY AS A GAUGE THEORY}
\end{center}
\vskip 0.3 cm

As we have seen, symmetry principles help in constructing the right classical
action. More importantly, they are often sufficient to ensure the viability of
a quantum theory obtained from the classical action. In particular, local or
gauge symmetry is the key to prove consistency (renormalizability) of the field
theories we know for the correct description of three of the four basic
interactions of nature. The gravitational interaction has stubbornly escaped
this rule in spite of the fact that, as we saw, it is described by a theory
based on general covariance, which is a local invariance quite analogous to
gauge symmetry. In this lecture we try to shed some light on this puzzle.

In 1955, less than a year after Yang and Mills proposed their model for
nonabelian gauge invariant interactions, Ryoyu Utiyama showed that the Einstein
theory can be written as a gauge theory for the Lorentz group \cite{Utiyama}.
This can be checked directly from the expression (\ref{Einstein1st}), which is
a Lorentz scalar and hence, trivially invariant under (local) Lorentz
transformations.

Our experience is that the manifold where we live is approximately flat, four
dimensional Minkowski spacetime. This space  is certainly invariant under the
Lorentz group $SO(3,1)$, but it also allows for translations. This means that
it would be nice to view $SO(3,1)$ as a subgroup of a larger group which
contains symmetries analogous to translations,
\begin{equation}
SO(3,1) \hookrightarrow G.  \label{Embedding}
\end{equation}
The smallest nontrivial choices for $G$ --which are not just $SO(3,1) \times
G_0$--, are:
\begin{equation}
G=\left\{
\begin{array}{cc}
SO(4,1) & \textrm{de Sitter} \\
SO(3,2) & \textrm{anti-de Sitter} \\
ISO(3,1) & \textrm{Poincar\'{e}}
\end{array}
\right.  \label{Lorentzembeddings}
\end{equation}
The de Sitter and anti-de Sitter groups are semisimple, while the Poincar\'{e}
group, which is a contraction of the other two, is not semisimple. (This is a
rather technical detail but it means that, unlike the Poincar\'{e} group, both
$SO(4,1)$ and $SO(3,2)$ are free of invariant abelian subgroups. Semisimple
groups are preferred as gauge groups because they have an invertible metric in
the group manifold.)

Since a general coordinate transformation
\begin{equation}
x^i \rightarrow x^i + \xi^i,  \label{Diffeo}
\end{equation}
looks like a local translation, it is natural to expect that diffeomorphism
invariance could be identified with the local boosts or translations necessary
to enlarge the Lorentz group into one of those close relatives in
(\ref{Lorentzembeddings}). Several attempts to carry out this identification,
however, have failed. The problem is that there seems to be no action for
general relativity, invariant under one of these extended groups
\cite{Kibble,Yang,Mansouri,MacDowell-Mansouri}. In other words, although the
fields $\omega^{ab}$ and $e^a$ match the generators of the group $G$, there is
no $G$-invariant 4-form available constructed with the building blocks listed
above.

As we shall see next, in odd dimensions ($D=2n-1$), and only in that case,
gravity can be cast as a gauge theory of the groups $SO(D,1)$, $SO(D-1,2)$, or
$ISO(D-1,1)$, in contrast with what one finds in dimension four.

\section{Lanczos-Lovelock Gravity}

We turn now to the construction of an action for gravity using the building
blocks at our disposal: $e^{a}$, $\omega _{\;b}^{a}$, $R_{\; b}^{a}$, $T^{a}$.
It is also allowed to include the only two invariant tensors of the Lorentz
group, $\eta _{ab}$, and $\epsilon _{a_{1}\cdot \cdot \cdot \cdot a_{D}}$ to
raise, lower and contract indices. The action must be an integral over the
$D$-dimensional spacetime manifold, which means that the lagrangian must be a
$D$-form. Since exterior forms are scalars under general coordinate
transformations, general covariance is guaranteed by construction and we need
not worry about it. The action principle cannot depend on the choice of basis
in the tangent space since Lorentz invariance should be respected. A sufficient
condition to ensure Lorentz invariance is to demand the lagrangian to be a
Lorentz scalar, although, as we will see, this is not strictly necessary.

Thus, we tentatively postulate the lagrangian for gravity to be a $D$-form
constructed by taking linear combinations of products of the above ingredients
in any possible way so as to form a Lorentz scalar. We exclude from the
ingredients functions such as the metric and its inverse, which rules out the
Hodge $\star $-dual. The only justification for this is that: {\bf i}) it
reproduces the known cases, and {\bf ii}) it explicitly excludes inverse
fields, like $e_{a}^{\mu }(x)$, which would be like $A_{\mu }^{-1}$ in
Yang-Mills theory (see \cite{Zumino} and \cite{Regge} for more on this). This
postulate rules out the possibility of including tensors like the Ricci tensor
$R_{\mu \nu }=\eta _{ac}e_{\mu }^{c}e_{b}^{\lambda }R_{\; \lambda \nu }^{ab}$,
or $R_{\alpha \beta }R_{\mu \nu }R^{\alpha \mu \beta \nu }$, etc. That this is
sufficient and necessary to account for all sensible theories of gravity in $D$
dimensions is the contents of a theorem due to David Lovelock \cite{Lovelock},
which in modern language can be stated thus:

\smallskip{\bf Theorem }[Lovelock,1970-Zumino,1986]: The most general action for
gravity that does not involve torsion, which gives at most second order field
equations for the metric and is of the form
\begin{equation}
I_{D}=\kappa \int \sum_{p=0}^{[D/2]}\alpha _{p}L^{(D,p)},
\label{LL-action}
\end{equation}
where the $\alpha_{p}$s are arbitrary constants, and $L^{(D,p)}$
is given by
\begin{equation}
L^{(D,\;p)}=\epsilon _{a_{1}\cdots a_{d}}R^{a_{1}a_{2}}\!\cdot
\!\cdot \!\cdot \!R^{a_{2p-1}a_{2p}}e^{a_{2p+1}}\!\cdot \!\cdot
\!\cdot \!e^{a_{D}}. \label{Lovlag}
\end{equation}
Here and in what follows we omit the wedge symbol in the exterior products. For
$D=2$ this action reduces to a linear combination of the $2$-dimensional Euler
character, $\chi _{2}$, and the spacetime volume (area),
\begin{eqnarray}
I_{2} &=&\kappa \int \alpha _{0}L^{(2,\;0)}+\alpha _{1}L^{(2,\; 1)}
\nonumber \\
&=&\kappa \int \sqrt{|g|}\left( \frac{\alpha _{0}}{2}R + 2\alpha_{1}\right)
d^{2}x  \label{2DGrav} \\
&=&\alpha _{0}^{\prime }\cdot \chi _{2}+\alpha _{1}^{\prime}\cdot V_{2}.
\nonumber
\end{eqnarray}
This action has only one local extremum, $V=0$, which reflects the fact that,
unless other matter sources are included, $I_{2}$ does not make a very
interesting dynamical theory for the geometry. If the geometry is restricted to
have a prescribed boundary this action describes the shape of a soap bubble,
the famous Plateau problem: {\em What is the surface of minimal area that has a
certain fixed closed curve as boundary?}.

For $D=3$, (\ref{LL-action}) reduces to the Hilbert action with cosmological
constant, and for $D=4$ the action picks up in addition the four dimensional
Euler invariant $\chi _{4}$. For higher dimensions the lagrangian is a
polynomial in the curvature 2-form of degree $d\leq D/2$. In even dimensions
the highest power in the curvature is the Euler character $\chi_{D}$. Each term
$L^{(D,\;p)}$ is the continuation to $D$ dimensions of the Euler density from
dimension $p<D$ \cite{Zumino}.

One can be easily convinced, assuming the torsion tensor vanishes identically,
that the action (\ref{LL-action}) is the most general scalar $D$-form that be
constructed using the building blocks we considered. The first nontrivial
generalization of Einstein gravity occurs in five dimensions, where a quadratic
term can be added to the lagrangian. In this case, the 5-form
\begin{equation}
\epsilon _{abcde}R^{ab}\!R^{cd}e^{e}\!= \sqrt{|g|}\left[R^{\alpha
\beta \gamma \delta }R_{\alpha \beta \gamma \delta }-4R^{\alpha
\beta }R_{\alpha \beta }+R^{2}\right] d^{5}x  \label{G-B}
\end{equation}
can be identified as the Gauss-Bonnet density, whose integral in four
dimensions gives the Euler character $\chi_{4}$. In 1938, Cornelius Lanczos
noticed that this term could be added to the Einstein-Hilbert action in five
dimensions \cite{Lanczos}. It is intriguing that he did not go beyond $D=5$.
The generalization to arbitrary $D$ was obtained by Lovelock more than 30 years
later as the Lanczos -Lovelock ({\bf LL}) lagrangians,
\begin{equation}
L_{D}=\sum_{p=0}^{[D/2]}\alpha _{p}L^{(D,p)}.  \label{LL-Lagrangian}
\end{equation}
These lagrangians were also identified as describing the only ghost-free
effective theories for spin two fields\footnote{Physical states in quantum
field theory have positive probability, which means that they are described by
positive norm vectors in a Hilbert space. Ghosts instead, are unphysical states
of negative norm. A lagrangian containing arbitrarily high derivatives of
fields generally leads to ghosts. Thus, the fact that a gravitational
lagrangian such as \ref{LL-Lagrangian} leads to a ghost-free theory is highly
nontrivial.}, generated from string theory at low energy
\cite{Zwiebach,Zumino}. From our perspective, the absence of ghosts is only a
reflection of the fact that the LL action yields at most second order field
equations for the metric, so that the propagators behave as $\alpha k^{-2}$,
and not as $\alpha k^{-2}+\beta k^{-4}$, as would be the case in a higher
derivative theory.

\subsection{Dynamical Content}

Extremizing the LL action (\ref{LL-action}) with respect to $e^{a}$ and
$\omega^{ab}$, yields
\begin{equation}
\delta I_{D}=\int [\delta e^{a}{\cal E}_{a}+\delta \omega^{ab}{\cal E}_{ab}]=0,
\label{Var-action}
\end{equation}
modulo surface terms. The condition for $I_{D}$ to have an extreme under
arbitrary first order variations is that the coefficients ${\cal E}_{a}$ ${\cal
E}_{ab}$ vanish identically. This condition is the geometry satisfies the field
equations
\begin{equation}
{\cal E}_{a}=\sum_{p=0}^{[\frac{D-1}{2}]}\alpha_{p}(d-2p){\cal E}_{a}^{(p)}=0,
\label{D-curvature}
\end{equation}
and
\begin{equation}
{\cal E}_{ab}=\sum_{p=1}^{[\frac{D-1}{2}]}\alpha_{p}p(d-2p){\cal
E}_{ab}^{(p)}=0, \label{D-torsion}
\end{equation}
where we have defined
\begin{eqnarray}
{\cal E}_{a}^{(p)}:= &&\epsilon_{ab_{2}\cdots b_{D-1}} R^{b_{2}b_{3}} \cdots
R^{b_{2p}b_{2p+1}}e^{b_{2p+1}}\cdots e^{b_{D}},\hspace{-0.06in}  \label{ELL1}
\\
{\cal E}_{ab}^{(p)}:= &&\epsilon_{aba_{3} \cdots a_{d}} R^{a_{3}a_{4}} \cdots
R^{a_{2p-1}a_{2p}} T^{a_{2p+1}} e^{a_{2p+2}} \cdots e^{a_{D}}.\hspace{-0.06in}
\label{ELL2}
\end{eqnarray}

These equations involve only first derivatives of $e^{a}$ and
$\omega_{\;b}^{a}$, simply because $d^{2}=0$. If one furthermore assumes, as is
usually done, that the torsion vanishes,
\begin{equation}
T^{a}=de^{a}+\omega _{\;b}^{a}e^{b}=0,  \label{0-Torsion}
\end{equation}
Eq. (\ref{ELL2}) is automatically satisfied and can be solved for $\omega$ as a
function of derivative of $e$ and its inverse $\omega =\omega (e,$ $\partial
e)$. Substituting the spin connection back into (\ref{ELL1}) yields second
order field equations for the metric. These equations are identical to the ones
obtained from varying the LL action written in terms of the Riemann tensor and
the metric,
\begin{equation}
I_{D}[g]=\int d^{D}x\sqrt{g}\left[ \alpha _{o}^{\prime }+ \alpha_{1}^{\prime}R
+ \alpha_{2}^{\prime }(R^{\alpha \beta \gamma \delta} R_{\alpha \beta \gamma
\delta}- 4R^{\alpha \beta }R_{\alpha \beta} + R^{2}) + \cdot \cdot \cdot
\right] . \label{LL-metric}
\end{equation}
This purely metric form of the action is the so-called second order formalism.
It might seem surprising that the action (\ref{LL-metric}) yields only second
order field equations for the metric, since the lagrangian contains second
derivatives of $g_{\mu \nu}$. In fact, it is sometimes asserted that the
presence of terms quadratic in curvature necessarily bring in higher order
equations for the metric but, as we have seen, this is not true for the LL
action. Higher derivatives of the metric would mean that the initial conditions
required to uniquely determine the time evolution are not those of General
Relativity and hence the theory would have different degrees of freedom from
standard gravity. It also means that the propagators in the quantum theory
develops poles at imaginary energies: ghosts. Ghost states spoil the unitarity
of the theory, making it hard to interpret its predictions.

One important feature that makes the LL theories very different for $D>4$ from
those for $D\leq 4$ is the fact that in the first case the equations are
nonlinear in the curvature tensor, while in the latter case all equations are
linear in $R^{ab}$ and in $T^{a}$. In particular, while for $D\leq 4$ the
equations (\ref{ELL2}) imply the vanishing of torsion, this is no longer true
for $D>4$. In fact, the field equations evaluated in some configurations may
leave some components of the curvature and torsion tensors indeterminate. For
example, Eq.(\ref{D-torsion}) has the form of a polynomial in $R^{ab}$ times
$T^{a}$, and it is possible that the polynomial vanishes identically, leaving
the torsion tensor completely arbitrary. However, the configurations for which
the equations do not determine $R^{ab}$ and $T^{a}$ completely form sets of
measure zero in the space of geometries. In a generic case, outside of these
degenerate configurations, the LL theory has the same $D(D-3)/2$ degrees of
freedom as in ordinary gravity \cite{Te-Z}.

\subsection{Adding Torsion}

Lovelock's theorem assumes torsion to be identically zero. If equation
(\ref{0-Torsion}) is assumed as an identity, means that $e^{a}$ and $\omega
_{\;b}^{a}$ are no longer independent fields, contradicting the assumption that
these fields correspond to two equally independent features of the geometry.
Moreover, for $D\leq 4$, equation (\ref{0-Torsion}) coincides with
(\ref{ELL2}), so that imposing the torsion-free constraint is, in the best
case, unnecessary.

On the other hand, if the field equation for a some field $\phi$ can be solved
algebraically as $\phi =f(\psi)$ in terms of the other fields, then by the
implicit function theorem, the reduced action principle $I[\phi,\psi]$ is
identical to the one obtained by substituting $f(\psi)$ in the action,
$I[f(\psi),\psi]$. This occurs in 3 and 4 dimensions, where the spin connection
can be algebraically obtained from its own field equation and $I[\omega
,e]=I[\omega (e,\partial e),e]$ . In higher dimensions, however, the
torsion-free condition is not necessarily a consequence of the field equations
and although (\ref{ELL2}) is algebraic in $\omega $, it is practically
impossible to solve for $\omega $ as a function of $e$. Therefore, it is not
clear in general whether the action $I[\omega ,e]$ is equivalent to the second
order form of the LL action, $I[\omega (e,\partial e),e]$.

Since the torsion-free condition cannot be always obtained from the field
equations, it is natural to look for a generalization of the Lanczos-Lovelock
action in which torsion is not assumed to vanish. This generalization consists
of adding of all possible Lorentz invariants involving $T^{a}$ explicitly (this
includes the combination $DT^{a}=R^{ab}e_{b}$). The general construction was
worked out in \cite{M-Z}. The main difference with the torsion-free case is
that now, together with the dimensional continuation of the Euler densities,
one encounters the Pontryagin (or Chern classes) as well.

For $D=3$, the only new torsion term not included in the Lovelock family is
\begin{equation}
e^{a}T_{a},  \label{eT}
\end{equation}
while for $D=4$, there are three terms not included in the LL series,
\begin{equation}
e^{a}e^{b}R_{ab} , T^{a}T_{a}  , R^{ab}R_{ab}.  \label{e2R+T2}
\end{equation}
The last term in (\ref{e2R+T2}) is the Pontryagin density, whose integral also
yields a topological invariant. It turns out that a linear combination of the
other two terms is also a topological invariant related to torsion known as the
Nieh-Yan density\cite{Nieh-Yan}
\begin{equation}
N_{4}=T^{a}T_{a}-e^{a}e^{b}R_{ab}.  \label{NY}
\end{equation}
The properly normalized integral of (\ref{NY}) over a 4-manifold is an integer
\cite{ChaZ}.

In general, the terms related to torsion that can be added to the action are
combinations of the form
\begin{eqnarray}
A_{2n} &=&e_{a_{1}}R_{\;a_{2}}^{a_{1}}R_{\;a_{3}}^{a_{2}}\cdot \cdot \cdot
R_{\;a_{n}}^{a_{n-1}}e^{a_{n}},\rm{ even} \: n\geq 2 \label{eRe} \\ B_{2n+1}
&=&T_{a_{1}}R_{\;a_{2}}^{a_{1}}R_{\;a_{3}}^{a_{2}}\cdot \cdot \cdot
R_{\;a_{n}}^{a_{n-1}}e^{a_{n}},\rm{ any } \: n\geq 1 \label{eRT} \\ C_{2n+2}
&=&T_{a_{1}}R_{\;a_{2}}^{a_{1}}R_{\;a_{3}}^{a_{2}}\cdot \cdot \cdot
R_{\;a_{n}}^{a_{n-1}}T^{a_{n}},\rm{ odd } \: n\geq 1 \label{TRT}
\end{eqnarray}
which are $2n$, $2n+1$ and $2n+2$ forms, respectively. These Lorentz invariants
belong to the same family with the Pontryagin densities or Chern classes,
\begin{equation}
P_{2n}=R_{\;a_{2}}^{a_{1}} R_{\;a_{3}}^{a_{2}}\cdot \cdot \cdot
R_{\;a_{1}}^{a_{n}},\rm{ even} \: n\geq 2.  \label{Pontryagin}
\end{equation}

The lagrangians that can be constructed now are much more varied and there is
no uniform expression that can be provided for all dimensions. For example, in
8 dimensions, in addition to the LL terms, one has all possible 8-form made by
taking products among the elements of the set
\{$A_{4},A_{8},B_{3},B_{5},B_{7},C_{4},C_{8},P_{4},P_{8}$\}. They are
\begin{equation}
(A_{4})^{2}, A_{8}, (B_{3}B_{5}), (A_{4}C_{4}), (C_{4})^{2},
C_{8}, (A_{4}P_{4}), (C_{4}P_{4}), (P_{4})^{2}, P_{8}.
\label{Torsional 8D}
\end{equation}

To make life even more complicated, there are some linear combinations of these
products which are topological densities. In 8 dimensions these are the
Pontryagin forms
\begin{eqnarray*}
P_{8} &=&R_{\;a_{2}}^{a_{1}}R_{\;a_{3}}^{a_{2}}\cdot \cdot \cdot
R_{\;a_{1}}^{a_{4}}, \\
(P_{4})^{2} &=&(R_{\;b}^{a}R_{\;a}^{b})^{2},
\end{eqnarray*}
which occur also in the absence of torsion, and generalizations of the Nieh-Yan
forms,
\begin{eqnarray*}
(N_{4})^{2} &=&(T^{a}T_{a}-e^{a}e^{b}R_{ab})^{2}, \\ N_{4}P_{4}
&=&(T^{a}T_{a}-e^{a}e^{b}R_{ab})(R_{\;d}^{c}R_{\; c}^{d}),
\end{eqnarray*}
etc. (for details and extensive discussions, see Ref.\cite{M-Z}).

\section{Selecting Sensible Theories}

Looking at these expressions one can easily get depressed. The lagrangians look
awkward, the number of terms in them grow wildly with the dimension\footnote{As
it is shown in \cite{M-Z}, the number of torsion-dependent terms grows as the
partitions of $D/4$, which is given by the Hardy-Ramanujan formula, $p(D/4)\sim
\frac{1}{\sqrt{3}D}\exp [\pi \sqrt{D/6}]$.}. This problem is not only an
aesthetic one. The coefficients in front of each term in the lagrangian is
arbitrary and dimensionful. This problem already occurs in 4 dimensions, where
the cosmological constant has dimensions of [length]$^{-4}$, and as evidenced
by the outstanding cosmological constant problem, there is no theoretical
argument to fix its value in order to compare with the observations.

There is another serious objection from the point of view of quantum mechanics.
Dimensionful parameters in the action are potentially dangerous because they
are likely to give rise to uncontrolled quantum corrections. This is what makes
ordinary gravity \smallskip nonrenormalizable in perturbation theory: In 4
dimensions, Newton's constant has dimensions of length squared, or inverse mass
squared, in natural units. This means that as the order in perturbation theory
increases, more powers of momentum will occur in the Feynman graphs, making its
divergences increasingly worse. Concurrently, the radiative corrections to
these bare parameters would require the introduction of infinitely many
counterterms into the action to render them finite\cite{'tHooft}. But an
illness that requires infinite amount of medication is also incurable.

The only safeguard against the threat of uncontrolled divergences in the
quantum theory is to have some symmetry principle that fixes the values of the
parameters in the action and limits the number of possible counterterms that
could be added to the lagrangian. Thus, if one could find a symmetry argument
to fix the independent parameters in the theory, these values will be
``protected'' by the symmetry. A good indication that this might happen would
be if the coupling constants are all dimensionless, as in Yang-Mills theory.

As we will see in odd dimensions there is a unique combination of terms in the
action that can give the theory an enlarged symmetry, and the resulting action
can be seen to depend on a unique constant that multiplies the action.
Moreover, this constant can be shown to be quantized by a argument similar to
Dirac's quantization of the product of magnetic and electric charge
\cite{QuantumG}.

\subsection{Extending the Lorentz Group}

The coefficients $\alpha _{p}$ in the LL lagrangian (\ref{LL-Lagrangian}) have
dimensions $l^{D-2p}$. This is because the canonical dimension of the vielbein
is $[e^{a}]=l^{1}$, while the Lorentz connection has dimensions that correspond
to a true gauge field, $[\omega^{ab}]=$ $l^{0}$. This reflects the fact that
gravity is naturally only a gauge theory for the Lorentz group, where the
vielbein plays the role of a matter field, which is not a connection field but
transforms as a vector under Lorentz rotations.

Three-dimensional gravity is an important exception to this statement, in which
case $e^{a}$ plays the role of a connection. Consider the simplest LL
lagrangian in 3 dimensions, the Einstein-Hilbert term
\begin{equation}
L_{3}=\epsilon_{abc}R^{ab}e^{c}.  \label{D=3}
\end{equation}
Under an infinitesimal Lorentz transformation with parameter
$\lambda_{\,\;b}^{a}$, $\omega ^{ab}$ transforms as
\begin{eqnarray}
\delta \omega_{\,\;b}^{a} &=&D\lambda_{\,\;b}^{a} \label{delta w}\\
&=&d\lambda_{\,\;b}^{a} + \omega _{\,\;c}^{a}\lambda_{\,\; b}^{c} -\omega
_{\,\;b}^{c}\lambda _{\,\;c}^{a}, \nonumber
\end{eqnarray}
while $e^{c}$, $R^{ab}$ and $\epsilon_{abc}$ transform as tensors,
\begin{eqnarray*}
\delta e^{a} &=&\lambda _{\,\;c}^{a}e^{c} \\ \delta R_{\,}^{ab}
&=&\lambda _{\,\;c}^{a}R^{cb}+\lambda _{\,\; c}^{b}R^{ac}, \\
\delta \epsilon _{abc} &=&\lambda _{\,\;a}^{d}\epsilon
_{dbc}+\lambda _{\,\;b}^{d}\epsilon _{adc}+\lambda
_{\,\;c}^{d}\epsilon _{abd}.
\end{eqnarray*}
Combining these relations the Lorentz invariance of $L_{3}$ can be shown
directly. What is unexpected is that one can view $e^{a}$ as a gauge connection
for the translation group. In fact, if under ``local translations'' in tangent
space, parametrized by $\lambda ^{a}$, the vielbein transforms as a connection,
\begin{eqnarray}
\delta e^{a} &=&D\lambda ^{a}  \nonumber \\
&=&d\lambda ^{a}+\omega _{\,\;b}^{a}\lambda ^{b},
\label{LocalTrans}
\end{eqnarray}
the lagrangian $L_{3}$ changes by a total derivative,
\begin{equation}
\delta L_{3}=d[\epsilon _{abc}R^{ab}\lambda ^{c}]. \label{dL3}
\end{equation}
Thus, the action changes by a surface term which can be dropped under standard
boundary conditions. This means that, in three dimensions, ordinary gravity can
be viewed as a gauge theory of the Poincar\'{e} group. We leave it as an
exercise to the reader to prove this. (Hint: use the infinitesimal
transformations $\delta e$ and $\delta \omega $ to compute the commutators of
the second variations to obtain the Lie algebra of the Poincar\'{e} group.)

The miracle also works in the presence of a cosmological constant $\Lambda =\mp
\frac{1}{6l^{2}}$. Now the lagrangian (\ref{LL-Lagrangian}) is
\begin{equation}
L_{3}^{AdS}=\epsilon _{abc}(R^{ab}e^{c}\pm \frac{1}{3l^{2}}e^{a}e^{b}e^{c}),
\label{L3AdS}
\end{equation}
and the action is invariant --modulo surface terms-- under the infinitesimal
transformations,
\begin{eqnarray}
\delta \omega _{\,\;}^{ab} &=&[d\lambda _{\;}^{ab}+ \omega_{\,\;c}^{a}
\lambda^{cb}+\omega_{\,\;c}^{b} \lambda_{\;}^{ac}] \mp
\frac{1}{l^{2}}[e^{a}\lambda ^{b}-\lambda ^{a}e^{b}] \label{dw-AdS} \\ \delta
e^{a} &=&[\lambda _{\;b}^{a}e^{b}]\;+\;[d\lambda ^{a}+\omega
_{\,\;b}^{a}\lambda ^{b}].  \label{de-AdS}
\end{eqnarray}
These transformations can be cast in a more suggestive way as
\begin{eqnarray*}
\delta \left[
\begin{array}{cc}
\omega _{\,\;}^{ab} & l^{-1}e^{a} \\
-l^{-1}e^{b} & 0
\end{array}
\right] &=&d\left[
\begin{array}{cc}
\lambda _{\;}^{ab} & l^{-1}\lambda ^{a} \\
-l^{-1}\lambda ^{b} & 0
\end{array}
\right] +\left[
\begin{array}{cc}
\omega _{\;c}^{a} & \pm l^{-1}e^{a} \\
-l^{-1}e_{c} & 0
\end{array}
\right] \left[
\begin{array}{cc}
\lambda ^{cb} & l^{-1}\lambda ^{c} \\
-l^{-1}\lambda ^{b} & 0
\end{array}
\right] \\
&&+\left[
\begin{array}{cc}
\omega _{\;c}^{b} & \pm l^{-1}e^{b} \\
-l^{-1}e_{c} & 0
\end{array}
\right] \left[
\begin{array}{cc}
\lambda ^{ac} & l^{-1}\lambda ^{a} \\
-l^{-1}\lambda ^{c} & 0
\end{array}
\right] .
\end{eqnarray*}
This can also be written as
\[
\delta W^{AB}=dW_{\;}^{AB}+W_{\,\;C}^{A}\Lambda ^{CB}+ W_{\;C}^{B}\Lambda
^{AC},
\]
where the 1-form $W^{AB}$ and the 0-form $\Lambda ^{AB}$ stand for the
combinations
\begin{eqnarray}
W_{\;}^{AB} &=&\left[
\begin{array}{cc}
\omega _{\,\;}^{ab} & l^{-1}e^{a} \\
-l^{-1}e^{b} & 0
\end{array}
\right]  \label{baticonn} \\
\Lambda ^{AB} &=&\left[
\begin{array}{cc}
\lambda _{\;}^{ab} & l^{-1}\lambda ^{a} \\
-l^{-1}\lambda ^{b} & 0
\end{array}
\right] ,  \label{batiparam}
\end{eqnarray}
where $a,b,..=1,2,..D,$ while $A,B,...=1,2,..,D+1$. Clearly, $W_{\; }^{AB}$
transforms as a connection and $\Lambda^{AB}$ can be identified as the
infinitesimal transformation parameters, but for which group? A clue comes from
the fact that $\Lambda ^{AB}=-\Lambda ^{BA}$. This immediately indicates that
the group is one that leaves invariant a symmetric, real bilinear form, so it
must be one of the $SO(r,s)$ family. The signs ($\pm $) in the transformation
above can be traced back to the sign of the cosmological constant. It is easy
to check that this structure fits well if indices are raised and lowered with
the metric
\begin{equation}
\Pi ^{AB}=\left[
\begin{array}{cc}
\eta _{\,\;}^{ab} & 0 \\
0 & \pm 1
\end{array}
\right] ,  \label{tangentAdS}
\end{equation}
so that, for example, $W_{\,\;B}^{A}=\Pi _{BC}W^{AC}$. Then, the covariant
derivative in the connection $W$ of this metric vanishes identically,
\begin{equation}
D_{W}\Pi ^{AB}=d\Pi ^{AB}+W_{\,\;C}^{A}\Pi ^{CB}+W_{\,\; C}^{B}\Pi ^{AC}=0.
\end{equation}
Since $\Pi ^{AB}$ is constant, this last expression implies $W^{AB}+W^{BA}=0$,
in exact analogy with what happens with the Lorentz connection,
$\omega^{ab}+\omega ^{ba}=0$, where $\omega ^{ab}=\eta ^{bc}\omega _{\;c}^{a}$.
Indeed, this is a very awkward way to discover that the 1-form $W_{\;}^{AB}$ is
actually a connection for the group which leaves invariant the metric $\Pi
^{AB}$. Here the two signs in $\Pi ^{AB}$ correspond to the de Sitter ($+$) and
anti-de Sitter\ ($-$) groups, respectively.

Observe that what we have found here is an explicit way to immerse the Lorentz
group into a larger one, in which the vielbein has been promoted to a component
of a larger connection, on the same footing as the Lorentz connection.

The Poincar\'{e} symmetry is obtained in the limit $l\rightarrow \infty $. In
that case, instead of (\ref{dw-AdS}, \ref{de-AdS}) one has
\begin{eqnarray}
\delta \omega _{\,\;}^{ab} &=&[d\lambda _{\;}^{ab}+ \omega_{\,\;c}^{a} \lambda
^{cb}+\omega_{\,\;c}^{b}\lambda_{\; }^{ac}] \label{dw-Poinc} \\ \delta e^{a}
&=&[\lambda _{\;b}^{a}e^{b}]\;+\;[d\lambda ^{a}+\omega _{\,\;b}^{a}\lambda
^{b}].  \label{de-Poinc}
\end{eqnarray}
In this limit, the representation in terms of $W$ becomes inadequate because
the metric $\Upsilon^{AB}$ becomes degenerate (noninvertible) and is not clear
how to raise and lower indices anymore.

\subsection{More Dimensions}

Everything that has been said about the embedding of the Lorentz group into the
(A)dS group, starting at equation (\ref{dw-AdS}) is not restricted to $D=3$
only and can be done in any $D$. In fact, it is always possible to embed the
Lorentz group in $D$ dimensions into the de-Sitter, or anti-de Sitter groups,
\begin{equation}
SO(D-1,1)\hookrightarrow \left\{
\begin{array}{cc}
SO(D,1), & \Pi ^{AB}=\rm{diag }(\eta _{\,\;}^{ab},+1) \\
SO(D-1,2), & \Pi ^{AB}=\rm{diag }(\eta _{\,\;}^{ab},-1)
\end{array}
.\right.  \label{embeddingAdS}
\end{equation}
with the corresponding Poincar\'{e} limit, which is the familiar symmetry group
of Minkowski space.
\begin{equation}
SO(D-1,1)\hookrightarrow ISO(D-1,1).  \label{embeddingPoinc}
\end{equation}

Then, the question naturally arises: can one find an action for gravity in
other dimensions which is also invariant, not just under the Lorentz group, but
under one of its extensions, $SO(D,1)$, $SO(D-1,2)$, $ISO(D-1,1)$? As we will
see now, the answer to this question is affirmative in odd dimensions. There is
always a action for $D=2n-1$, invariant under local $SO(2n-2,2)$, $SO(2n-1,1)$
or $ISO(2n-2,1)$ transformations, in which the vielbein and the spin connection
combine to form the connection of the larger group. In even dimensions,
however, this cannot be done.

Why is it possible in three dimensions to enlarge the symmetry from local
$SO(2,1)$ to local $SO(3,1)$, $SO(2,2)$, $ISO(2,1)$? What happens if one tries
to do this in four or more dimensions? Let us start with the Poincar\'{e} group
and the Hilbert action for $D=4$,
\begin{equation}
L_{4}=\epsilon _{abcd}R^{ab}e^{c}e^{d}.  \label{D=4}
\end{equation}
Why is this not invariant under local translations $\delta e^{a}=
d\lambda^{a}+\omega _{\,\;b}^{a}\lambda ^{b}$? A simple calculation yields
\begin{eqnarray}
\delta L_{4} &=&2\epsilon _{abcd}R^{ab}e^{c}\delta e^{d}  \nonumber \\
&=&d(2\epsilon _{abcd}R^{ab}e^{c}\lambda ^{d})+2\epsilon
_{abcd}R^{ab}T^{c}\lambda ^{d}.  \label{delta4}
\end{eqnarray}
The first term in the r.h.s. of (\ref{delta4}) is a total derivative and
therefore gives a surface contribution to the action. The last term, however,
need not vanish, unless one imposes the field equation $T^{a}=0$. But this
means that the invariance of the action only occurs on shell. On shell
symmetries are not real symmetries and they need not survive quantization. On
close inspection, one observes that the miracle occurred in 3 dimensions
because the lagrangian contained only one $e$. This means that a lagrangian of
the form
\begin{equation}
L_{2n+1}=\epsilon _{a_{1}\cdot \cdot \cdot \cdot
a_{2n+1}}R^{a_{1}a_{2}}\cdot \cdot \cdot R^{a_{2n-1}a_{2n}}e^{a_{2n+1}}
\label{odd-D-Poinc}
\end{equation}
is invariant under local Poincar\'{e} transformations (\ref{dw-Poinc},
\ref{de-Poinc}), as can be easily checked out. Since the Poincar\'{e} group is
a limit of (A)\smallskip dS, it seem likely that there should exist a
lagrangian in odd dimensions, invariant under local (A)dS transformations,
whose limit for $l\rightarrow \infty $ (vanishing cosmological constant) is
(\ref{odd-D-Poinc}). One way to find out what that lagrangian might be, one
could take the most general LL lagrangian and select the coefficients by
requiring invariance under (\ref{dw-AdS}, \ref{de-AdS}). This is a long,
tedious and sure route. An alternative approach is to try to understand why it
is that in three dimensions the gravitational lagrangian with cosmological
constant (\ref{L3AdS}) is invariant under the (A)dS group.

If one takes seriously the notion that $W^{AB}$ is a connection, then one can
compute the associated curvature,
\[
F^{AB}=dW^{AB}+W_{\;C}^{A}W^{CB},
\]
using the definition of $W^{AB}$ (\ref{baticonn}). It is a simple exercise to
prove
\begin{equation}
F_{\;}^{AB}=\left[
\begin{array}{cc}
R_{\,\;}^{ab}\pm l^{-2}e^{a}e^{b} & l^{-1}T^{a} \\
-l^{-1}T^{b} & 0
\end{array}
\right] .  \label{baticurvature}
\end{equation}
If $a,b$ run from 1 to 3 and $A,B$ from 1 to 4, then one can construct the
4-form invariant under the (A)dS group,
\begin{equation}
E_{4}=\epsilon _{ABCD}F_{\;}^{AB}F_{\;}^{CD},  \label{E=F2}
\end{equation}
which is readily recognized as the Euler density in a four-dimensional manifold
whose tangent space is not Minkowski, but has the metric $\Pi ^{AB}= $diag
$(\eta _{\,\;}^{ab},\mp 1)$. $E_{4}$ can also be written explicitly in terms of
$R^{ab}$, $T^{a}$, and $e^{a}$,
\begin{eqnarray}
E_{4} &=&4\epsilon _{abc}(R_{\,\;}^{ab}\pm l^{-2}e^{a}e^{b})l^{-1}T^{a} \label{E4} \\
&=&\frac{4}{l}d\left[ \epsilon _{abc}\left( R_{\,\;}^{ab}\pm \frac{1}{ 3l^{2}}e^{a}e^{b}\right)
e^{a}\right] ,  \nonumber
\end{eqnarray}
which is, up to constant factors, the exterior derivative of the
three-dimensional lagrangian (\ref{L3AdS}),
\begin{equation}
E_{4}=\frac{4}{l}dL_{3}^{AdS}.  \label{E4=dL3}
\end{equation}

This explains why the action is (A)dS invariant up to surface terms: the l.h.s.
of (\ref{E4=dL3}) is invariant by construction under local (A)dS, so the same
must be true of the r.h.s., $\delta \left( dL_{3}^{AdS}\right) =0$. Since the
variation ($\delta $) is a linear operation,
\[
d\left( \delta L_{3}^{AdS}\right) =0,
\]
which in turn means, by Poincar\'{e}'s Lemma that, locally, $\delta
L_{3}^{AdS}= d(something)$. That is exactly what we found for the variation,
[see, (\ref{dL3})]. The fact that three dimensional gravity can be written in
this way was observed many years ago in Refs. \cite{Achucarro-Townsend,Witten}.

The key to generalize the (A)\smallskip dS lagrangian from 3 to $2n-1$
dimensions is now clear\footnote{The construction we outline here was discussed
by Chamseddine \cite{Chamseddine}, M\"{u}ller-Hoissen \cite{Muller-Hoissen},
and Ba\~{n}ados, Teitelboim and this author in \cite{BTZ94}.}. First,
generalize the Euler density (\ref{E=F2}) to a $2n$-form,
\begin{equation}
E_{2n}=\epsilon _{A_{1}\cdot \cdot \cdot A_{2n}}F^{A_{1}A_{2}}\cdot \cdot
\cdot F^{A_{2n-1}A_{2n}}.  \label{E2n=Fn}
\end{equation}

Second, express $E_{2n}$ explicitly in terms of $R^{ab}$, $T^{a}$, and $e^{a}$,
and write this as the exterior derivative of a $(2n-1)$-form which can be used
as a lagrangian in $(2n-1)$ dimensions. Direct computation yields the $(2n-1)$-
dimensional lagrangian as
\begin{equation}
L_{2n-1}^{(A)dS}=\sum_{p=0}^{n-1}\bar{\alpha}_{p}L^{(2n-1,p)},
\label{(A)dS2n+1}
\end{equation}
where $L^{(D,p)}$ is given by (\ref{Lovlag}) and the coefficients
$\bar{\alpha}_{p}$ are no longer arbitrary, but they take the values
\begin{equation}
\smallskip \bar{\alpha}_{p}=\kappa \cdot \frac{(\pm 1)^{p+1}l^{2p-D}}{(D-2p)}
\left(
\begin{array}{c}
n-1 \\
p
\end{array}
\right) ,\;p=1,2,...,n-1=\frac{D-1}{2},  \label{alphasCS}
\end{equation}
where $\kappa$ is an arbitrary dimensionless constant. It is left as an
exercise to the reader to check that $dL_{2n+1}^{(A)dS} = E_{2n}$ and to show
the invarianceof $L_{2n-1}^{(A)dS}$ under the (A)dS group. In five dimensions,
for example, the (A)dS lagrangian reads
\begin{equation}
L_{5}^{(A)dS}=\kappa \cdot \epsilon _{abcde}\left[ \frac{1}{l} e^{a}R^{bc}
R^{de}\pm \frac{2}{3l^{3}}e^{a}e^{b}e^{c}R^{de}+\frac{1}{5l^{5}} e^{a} e^{b}
e^{c} e^{d}e^{e}\right] . \label{(A)dS5}
\end{equation}

The parameter $l$ is a length scale --the Planck length-- and cannot be fixed
by other considerations. Actually, $l$ only appears in the combination
\[
\tilde{e}^{a}=\frac{e^{a}}{l},
\]
which could be considered as the ``true'' dynamical field, which is the natural
thing to do if one uses $W^{AB}$ instead of $\omega ^{ab}$ and $e^{a} $
separately. In fact, the lagrangian (\ref{(A)dS2n+1}) can also be written in
terms of $W^{AB}$ and its exterior derivative, as
\begin{equation}
L_{2n-1}^{(A)dS}=\kappa \cdot \epsilon _{A_{1}\cdot \cdot \cdot \cdot
A_{2n}}\left[ W(dW)^{n-1}+a_{3}W^{3}(dW)^{n-2}+\cdot \cdot \cdot
a_{2n-1}W^{2n-1}\right] ,  \label{(A)dS2n+1'}
\end{equation}
where all indices are contracted appropriately and the coefficients $a_{3}$,
$a_{5}$, are all combinatoric factors without dimensions.

The only remaining free parameter is $\kappa$. Suppose this lagrangian is used
to describe a simply connected, compact $2n-1$ dimensional manifold $M$, which
is the boundary of a $2n$-dimensional compact orientable manifold $\Omega $.
Then the action for the geometry of $M$ can be expressed as the integral of the
Euler density $E_{2n}$ over $\Omega $, multiplied by $\kappa$. But since there
can be many different manifolds with the same boundary $M$, the integral over
$\Omega $ should give the physical predictions as that over another manifold,
$\Omega ^{\prime }$. In order for this change to leave the path integral
unchanged, a minimal requirement would be
\begin{equation}
\kappa \left[ \int_{\Omega }E_{2n}-\int_{\Omega^{\prime
}}E_{2n}\right] =2n\pi \hbar .  \label{quantum}
\end{equation}
The quantity in brackets --with right normalization-- is the Euler number of
the manifold obtained by gluing $\Omega $ and $\Omega ^{\prime}$ along $M$, in
the right way to produce an orientable manifold, $\chi [\Omega \cup \Omega
^{\prime }\dot{]}$, which can take an arbitrary integer value. From this, one
concludes that $\kappa $ must be quantized\cite{QuantumG},
\[
\kappa =nh.
\]
where $h$ is\ Planck's constant.

\subsection{Chern-Simons}

There is a more general way to look at these lagrangians in odd dimensions,
which also sheds some light on their remarkable enlarged symmetry. This is
summarized in the following

{\bf Lemma:} Let $C(F)$ be an invariant $2n$-form constructed with the field
strength $F=dA+A^{2}$, where $A$ is the connection for some gauge group $G$. If
there exists a $2n-1$ form, $L$, depending on $A$ and $dA$, such that $dL=C$,
then under a gauge transformation, $L$ changes by a total derivative (exact
form).

The $(2n-1)-$form $L$ is known as the Chern-Simons ({\bf CS}) lagrangian. This
lemma shows that $L$ defines a nontrivial lagrangian for $A$ which{\em is not
invariant} under gauge transformations, but that changes by a function that
only depends on the fields at the boundary.

This construction is not only restricted to the Euler invariant discussed
above, but applies to any invariant of similar nature, generally known as
characteristic classes. Other well known characteristic classes are the
Pontryagin or Chern classes and their corresponding CS forms were studied first
in the context of abelian and nonabelian gauge theories (see, e. g.,
\cite{Jackiw,Nakahara}).

The following table gives examples of CS forms which define lagrangians in
three dimensions, and their corresponding characteristic classes, \vskip 0.3cm
\begin{center}
\begin{tabular}{|c|c|c|}
\hline $\rm{Lagrangian}$ & $ \rm{CS ~form }L$ & $dL$ \\ \hline
\hline $L_{3}^{Lor}$ & $\omega _{\;b}^{a}d\omega
_{\;a}^{b}+\frac{2}{3} \omega _{\;b}^{a}\omega _{\;c}^{b}\omega
_{\;a}^{c}$ & $R_{ \;b}^{a}R_{\;a}^{b}$ \\ \hline $L_{3}^{Tor}$ &
$e^{a}T_{a}$ & $T^{a}T_{a}-e^{a}e^{b}R_{ab}$ \\ \hline
$L_{3}^{U(1)}$ & $AdA$ & $FF$ \\ \hline $L_{3}^{SU(N)}$ & $tr[{\bf
A}d{\bf A+}\frac{2}{3}{\bf AAA]}$ & $tr[{\bf FF}]$ \\ \hline
\end{tabular}
\end{center}
\vskip 0.3cm

In this table, $R$, $F$, and ${\bf F}$ are the curvatures of the Lorentz
connection $\omega _{\;b}^{a}$ , the electromagnetic ($U(1)$) connection $A$,
and the Yang-Mills ($SU(N)$) connection {\bf A}, respectively.

\subsection{Torsional CS}

So far we have not included torsion in the CS lagrangians, but as we see in the
third row of the table above it is also possible to construct CS forms that
include torsion. All the CS forms above are Lorentz invariant up to a closed
form, but there is a linear combination of the first two which is invariant
under the (A)dS group. The so-called exotic gravity, given by
\begin{equation}
L_{3}^{Exotic}=L_{3}^{Lor}+\frac{2}{l^{2}}L_{3}^{Tor},  \label{3d-exotic}
\end{equation}
is invariant under (A)dS, as can be shown by computing its exterior derivative,
\begin{eqnarray*}
dL_{3}^{Exotic} &=&R_{\;b}^{a}R_{\;a}^{b}\pm \frac{2}{l^{2}} \left(
T^{a}T_{a}-e^{a}e^{b}R_{ab}\right) \\ &=&F_{\;B}^{A}F_{\;A}^{B}.
\end{eqnarray*}
This exotic lagrangian has the curious property of giving exactly the same
field equations as the standard $dL_{3}^{AdS}$, but interchanged: the equation
for $e^{a}$ form one is the equation for $\omega ^{ab}$ of the other. In five
dimensions there are no new terms due to torsion, and in seven there are three
torsional CS terms,
\[
\begin{array}{|c|c|c|}
\hline \rm{Lagrangian} & \rm{CS ~form }L & dL \\ \hline \hline
L_{7}^{Lor} & \omega (d\omega)^{3} + \cdot \cdot \cdot + \frac{4}{7}
\omega ^{7} & R_{\;b}^{a}R_{\;c}^{b}R_{\;d}^{c}R_{\;a}^{d} \\
\hline L_{3}^{Lor}R_{\;b}^{a}R_{\;a}^{b} & (\omega_{\; b}^{a}
d\omega_{\;a}^{b}+\frac{2}{3}\omega _{\;b}^{a}\omega_{ \;c}^{b}
\omega_{\;a}^{c})R_{\;b}^{a}R_{\;a}^{b} & \left( R_{\;b}^{a} R_{\;a}^{b}\right)
^{2} \\ \hline L_{3}^{Tor}R_{\;b}^{a}R_{\;a}^{b} & e^{a}T_{a}R_{\;
b}^{a}R_{\;a}^{b} & (T^{a}T_{a}-e^{a}e^{b}R_{ab})R_{\;b}^{a}R_{ \;a}^{b}\\
\hline
\end{array}
\]
\vskip 0.3cm

In three spacetime dimensions, GR is a renormalizable quantum theory
\cite{Witten}. It is strongly suggestive that precisely in 2+1 dimensions this
is also a gauge theory on a fiber bundle. It could be thought that the exact
solvability miracle is due to the absence of propagating degrees of freedom in
three-dimensional gravity, but the final power-counting argument of
renormalizability rests on the fiber bundle structure of the Chern-Simons
system and doesn't seem to depend on the absence of propagating degrees of
freedom.

\subsection{Even Dimensions}

The CS construction fails in $2n$ dimensions for the simple reason that there
are no characteristic classes $C(F)$ constructed with products of curvature in
$2n+1$ dimensions. This is why an action for gravity in even dimensions cannot
be invariant under the (anti-) de Sitter or Poincar\'{e} groups. In this light,
it is fairly obvious that although ordinary Einstein-Hilbert gravity can be
given a fiber bundle structure for the Lorentz group, this structure cannot be
extended to include local translational invariance.

In some sense, the closest one can get to a CS theory in even dimensions is the
so-called Born-Infeld ({\bf BI}) theories \cite {JJG,BTZ94,Tr-Z}. The BI
lagrangian is obtained by a particular choice of the $\alpha _{p}$s in the LL
series, so that the lagrangian takes the form
\begin{equation}
L_{2n}^{BI}=\epsilon _{a_{1}\cdot \cdot \cdot a_{2n}}\bar{R}^{a_{1}a_{2}}\cdot
\cdot \cdot \bar{R}^{a_{2n-1}a_{2n}},  \label{2nBI}
\end{equation}
where $\bar{R}^{ab}$ stands for the combination
\begin{equation}
\bar{R}^{ab}=R^{ab}\pm \frac{1}{l^{2}}e^{a}e^{b}.
\label{concircular}
\end{equation}

With this definition it is clear that the lagrangian (\ref{2nBI}) contains only
one free parameter, $l$. This lagrangian has a number of interesting classical
features like simple equations, black hole solutions, cosmological models, etc.
The simplification comes about because the equations admit a unique maximally
symmetric configuration given by $\bar{R}^{ab}=0$, in contrast with the
situation when all $\alpha _{p}$s are arbitrary. As we have mentioned, for
arbitrary $\alpha _{p}$s, the field equations do not determine completely the
components of $R^{ab}$ and $T^{a}$ in general. This is because the high
nonlinearity of the equations can give rise to degeneracies. The BI choice is
in this respect the best behaved since the degeneracies are restricted to only
one value of the radius of curvature ($R^{ab}\pm \frac{1}{l^{2}} e^{a} e^{b} =
0$). At the same time, the BI action has the least number of algebraic
constrains required by consistency among the field equations, and it is
therefore the one with the simplest dynamical behavior\cite{Tr-Z}.

Equipped with the tools to construct gravity actions invariant under larger
groups, in the next lecture we undertake the extension of this trick to include
supersymmetry.

\vskip 0.7cm

\begin{center}
{\bf LECTURE 3} \vskip 0.3 cm {\bf CHERN SIMONS SUPERGRAVITY}
\end{center}
\vskip 0.3cm

The previous lectures dealt with the possible ways in which pure gravity can be
extended by relaxing three standard assumptions of General Relativity: {\bf i})
that the notion of parallelism is derived from metricity, {\bf ii}) that the
dimension of spacetime must be four, and {\bf iii)} that the action should only
contain the Einstein Hilbert term $\sqrt{g}R$. On the other hand, we still
demanded that {\bf iv) }the metric components obey second order field
equations, {\bf v)} the lagrangian be an $D$-form constructed out of the
vielbein, $e^{a}$, the spin connection, $\omega _{\;b}^{a},$ and their exterior
derivatives, {\bf vi)} the action be invariant under local Lorentz rotations in
the tangent space. This allowed for the inclusion of several terms containing
higher powers of the curvature and torsion multiplied by arbitrary and
dimensionful coefficients. The presence of these arbitrary constants was
regarded as a bit of an embarrassment which could be cured by enlarging the
symmetry group, thereby fixing all parameters in the lagrangian and making the
theory gauge invariant under the larger symmetry group. The cure works in odd
but not in even dimensions. The result was a highly nonlinear Chern-Simons
theory of gravity, invariant under local AdS transformations in the tangent
space. We now turn to the problem of enlarging the contents of the theory to
allow for supersymmetry.

\section{Supersymmetry}

Supersymmetry is a symmetry most theoreticians are willing to accept as a
legitimate feature of nature, although it has never been experimentally
observed. The reason is that it is such a unique and beautiful idea that it is
commonly felt that it would be a pity if it is not somehow realized in nature.
Supersymmetry is the only symmetry which can accommodate spacetime and internal
symmetries in a nontrivial way. By nontrivial we mean that the Lie algebra {\em
is not} a direct sum of the algebras of spacetime and internal symmetries.
There is a famous no-go theorem which states that it is impossible to do this
with an ordinary Lie group, closed under commutator (antisymmetric product,
$\left[ \cdot ,\cdot \right] $). The way supersymmetry circumvents this
obstacle is by having both commutators and anticommutators (symmetric product,
$\left\{ \cdot ,\cdot \right\} $), forming what is known as a {\bf graded Lie
algebra}, also called a super Lie algebra or simply, a superalgebra. For a
general introduction to supersymmetry, see \cite{Sohnius, Freund}.

The importance of this unification is that it combines bosons and fermions on
the same footing. Bosons are the carriers of interactions, such as the photon,
the graviton and gluons, while fermions are the constituents of matter, such as
electrons and quarks. Thus, supersymmetry predicts the existence of a fermionic
carriers of interaction and bosonic constituents of matter as partners of the
known particles, none of which have been observed.

Supersymmetry also strongly restricts the possible theories of nature and in
some cases it even predicts the dimension of spacetime, like in superstring
theory as seen in the lectures by Stefan Theisen in this same volume \cite
{Theisen}.

\subsection{Superalgebra}

A superalgebra has two types of generators: bosonic, ${\bf B}_{i}$, and
fermionic, ${\bf F}_{\alpha }$. They are closed under the (anti-) commutator
operation, which follows the general pattern
\begin{eqnarray}
\left[ {\bf B}_{i}{\bf ,B}_{j}\right] &=&C_{ij}^{k}{\bf B}_{k}  \label{bb} \\
\left[ {\bf B}_{i}{\bf ,F}_{\alpha }\right] &=&C_{i\alpha }^{\beta
}{\bf F}_{\beta }  \label{bf} \\
\left\{ {\bf F}_{\alpha },{\bf F}_{\beta }\right\} &=&C_{\alpha \beta
}^{\gamma }{\bf B}_{\gamma }  \label{ff}
\end{eqnarray}
The generators of the Poincar\'{e} group are included in the bosonic sector,
and the ${\bf F}_{\alpha }$'s are the supersymmetry generators. This algebra,
however, does not close for an arbitrary bosonic group. In other words, given a
Lie group with a set of bosonic generators, it is not always possible to find a
set of fermionic generators to enlarge the algebra into a closed superalgebra.
The operators satisfying relations of the form (\ref{bb}-\ref{ff}), are still
required to satisfy a consistency condition, the super-Jacobi identity,
\begin{equation}
\lbrack {\bf G}_{\mu },[{\bf G}_{\nu }, {\bf G}_{\lambda }]_{\pm
}]_{\pm }+(-)^{\sigma (\nu \lambda \mu )}[{\bf G}_{\nu },[{\bf
G}_{\lambda },{\bf G}_{\mu }]_{\pm }]_{\pm }+(-)^{\sigma (\lambda
\mu \nu )}[{\bf G}_{\lambda },[{\bf G}_{\mu },{\bf G}_{\nu }]_{\pm
}]_{\pm }=0.  \label{superjacobi}
\end{equation}
Here ${\bf G}_{\mu }$ represents any generator in the algebra,
$[R,S]_{\pm}=RS\pm SR$, where this sign is chosen according the bosonic or
fermionic nature of the operators in the bracket, and $\sigma (\nu \lambda \mu
)$ is the number of permutations of fermionic generators.

As we said, starting with a set of bosonic operators it is not always possible
to find a set of $N$ fermionic ones that generate a closed superalgebra. It is
often the case that extra bosonic generators are needed to close the algebra,
and this usually works for some values of $N$ only. In other cases there is
simply no supersymmetric extension at all. This happens, for example, with the
de Sitter group, which has no supersymmetric extension in general
\cite{Freund}. For this reason in what follows we will restrict to AdS
theories.

\subsection{Supergravity}

The name supergravity ({\bf SUGRA}) applies to any of a number of
supersymmetric theories that include gravity in their bosonic sectors. The
invention/discovery of supergravity in the mid 70's came about with the
spectacular announcement that some ultraviolet divergent graphs in pure gravity
were cancelled by the inclusion of their supersymmetric partners \cite{PvN}.
For some time it was hoped that the nonrenormalizability of GR could be cured
in this way by its supersymmetric extension. However, the initial hopes raised
by SUGRA as a way taming the ultraviolet divergences of pure gravity eventually
vanished with the realization that SUGRAs would be nonrenormalizable as well
\cite{Townsend}.

Again, one can see that the standard form of SUGRA is not a gauge theory for a
group or a supergroup, and that the local (super-) symmetry algebra closes
naturally on shell only. The algebra could be made to close off shell by force,
at the cost of introducing auxiliary fields --which are not guaranteed to exist
for all $d$ and $N$ \cite{RT}--, and still the theory would not have a fiber
bundle structure since the base manifold is identified with part of the fiber.
Whether it is the lack of fiber bundle structure the ultimate reason for the
nonrenormalizability of gravity remains to be proven. It is certainly true,
however, that if GR could be formulated as a gauge theory, the chances for its
renormalizability would clearly increase. At any rate, now most high energy
physicists view supergravity as an effective theory obtained from string theory
in some limit. In string theory, eleven dimensional supergravity is seen as an
effective theory obtained from ten dimensional string theory at strong coupling
\cite{Theisen}. In this sense supergravity would not be a fundamental theory
and therefore there is no reason to expect that it should be renormalizable.

In any case, our point of view here is that there can be more than one system
that can be called supergravity, whose connection with the standard theory is
still not clear. As we have seen in the previous lecture, the CS gravitation
theories in odd dimensions are genuine (off-shell) gauge theories for the
anti-de Sitter ({\bf A)dS} or Poincar\'{e} groups.

\subsection{From Rigid Supersymmetry to Supergravity}

Rigid or global SUSY is a supersymmetry in which the group parameters are
constants throughout spacetime. In particle physics the spacetime is usually
assumed to have fixed Minkowski geometry. Then the relevant SUSY is the
supersymmetric extension of the Poincar\'{e} algebra in which the supercharges
are ``square roots'' of the generators of spacetime translations, $\{{\bf
\bar{Q}},{\bf Q}\}\sim \Gamma \cdot {\bf P}$. The extension of this to a local
symmetry can be done by substituting the momentum ${\bf P}_{\mu }=i\partial
_{\mu }$ by the generators of spacetime diffeomorphisms,{\bf \ }${\Bbb H}_{\mu
}$, and relating them to the supercharges by $\{{\bf \bar{Q}},{\bf Q}\}\sim
\Gamma \cdot {\Bbb H}$. The resulting theory has a local supersymmetry algebra
which only closes on-shell \cite{PvN}. As we discussed above, the problem with
on-shell symmetries is that they are not likely to survive in the quantum
theory.

Here we consider the alternative~approach of extending the AdS symmetry on the
tangent space into a supersymmetry rather than working directly on the
spacetime manifold. This point of view is natural if one recalls that spinors
are naturally defined relative to a local frame on the tangent space rather
than to the coordinate basis. In fact, spinors provide an irreducible
representation for $SO(N)$, but not for $GL(N)$, which describe infinitesimal
general coordinate transformations. The basic strategy is to reproduce the 2+1
``miracle'' in higher dimensions. This idea was applied in five
dimensions\cite{Chamseddine}, as well as in higher dimensions \cite
{BTrZ,TrZ1,TrZ2}.

\subsection{Assumptions of Standard Supergravity}

Three implicit assumptions are usually made in the construction of standard
SUGRA:

{\bf (i)} The fermionic and bosonic fields in the Lagrangian should come in
combinations such that they have equal number of propagating degrees of
freedom. This is usually achieved by adding to the graviton and the gravitini a
number of fields of spins $0,1/2$ and $1$ \cite{PvN}. This matching, however,
is not necessarily true in AdS space, nor in Minkowski space if a different
representation of the Poincar\'{e} group (e.g., the adjoint representation) is
used \cite{Sohnius}.

The other two assumptions concern the purely gravitational sector and are
dictated by economy:

{\bf (ii)} gravitons are described by the Hilbert action (plus a possible
cosmological constant), and,

{\bf (iii)} the spin connection and the vielbein are not independent fields but
are related through the torsion equation.

The fact that the supergravity generators do not form a closed off-shell
algebra can be traced back to these assumptions.

The argument behind {\bf (i)} is closely related to the idea that the fields
should be in a vector representation of the Poincar\'{e} group. This assumption
comes from the interpretation of supersymmetric states as represented by the
in- and out- plane waves in an asymptotically free, weakly interacting theory
in a Minkowski background. Then, because the hamiltonian commutes with the
supersymmetry generators, every nonzero mass state must have equal number of
bosonic and fermionic states: For each bosonic state of energy, $|E>_{B}$,
there is a fermionic one with the same energy, $|E>_{F}={\bf Q}$ $|E>_{B}$, and
vice versa. This argument, however, breaks down if the Poincar\'{e} group in
not a symmetry of the theory, as it happens in an asymptotically AdS space, and
in other simple cases such as SUSY in 1+1, with broken translational invariance
\cite{LSV}.

Also implicit in the argument for counting the degrees of freedom is the usual
assumption that the kinetic terms and couplings are those of a minimally
coupled gauge theory, a condition that is not met by a CS theory. Apart from
the difference in background, which requires a careful treatment of the unitary
irreducible representations of the asymptotic symmetries \cite {Fronsdal}, the
counting of degrees of freedom in CS theories is completely different from the
counting for the same connection 1-forms in a YM theory (see Lecture 4 below).

\section{Super AdS algebras}

In order to construct a supergravity theory that contains gravity with a
cosmological constant, a mathematically oriented physicist would look for the
smallest superalgebra that contains the generators of the AdS algebra. This was
asked --and answered!-- many years ago, at least for some dimensions $D=2,3,4$
$mod$ $8$, \cite{vH-VP}. However this is not all, we would also want to see an
action that realizes the symmetry. Constructing a supergravity action for a
given dimension that includes a cosmological constant is a nontrivial task. For
example, the standard supergravity in eleven dimensions has been know for a
long time \cite{CJS}, however, it does not contain a cosmological constant
term, and it has been shown to be impossible to accommodate one \cite{BDHS}.
Moreover, although it was known to the authors of Ref.\cite{CJS} that the
supergroup that contains the AdS group in eleven dimensions is $SO(32|1)$, no
action was found for almost twenty years for the theory of gravity which
exhibits this symmetry.

An explicit representation of the superalgebras that contain AdS algebra
$so(D-1,2)$ can be constructed along the lines of \cite{vH-VP}, although here
we consider an extension of this method which applies to the cases $D=5,$ $7, $
and $9$ as well \cite{TrZ1}. The crucial observation is that the Dirac matrices
provide a natural representation of the AdS algebra in any dimension. Then, the
AdS connection ${\bf W}$ can be written in this representation as ${\bf
W}=e^{a}{\bf J}_{a}+\frac{1}{2}\omega ^{ab}{\bf J}_{ab}$, where

\begin{equation}
{\bf J}_{a}=\left[
\begin{array}{cc}
\frac{1}{2}(\Gamma _{a})_{\beta }^{\alpha } & 0 \\
0 & 0
\end{array}
\right] ,  \label{ja}
\end{equation}

\begin{equation}
{\bf J}_{ab}=\left[
\begin{array}{cc}
\frac{1}{2}(\Gamma _{ab})_{\beta }^{\alpha } & 0 \\
0 & 0
\end{array}
\right] .  \label{jab}
\end{equation}
Here ${\bf \Gamma }_{a}$, $a=1,...,D$ are $m\times m$ Dirac matrices, where
$m=2^{[D/2]}$ (here $[r]$ denotes the integer part of $r$), and ${\bf \Gamma
}_{ab}=\frac{1}{2}[{\bf \Gamma }_{a},{\bf \Gamma }_{b}]$. These two class of
matrices form a closed commutator subalgebra (the AdS algebra) of the {\bf
Dirac algebra} ${\cal D}$, obtained by taking antisymmetrized products of ${\bf
\Gamma}$ matrices
\begin{equation}
{\bf I},{\bf \Gamma }_{a},{\bf \Gamma }_{a_{1}a_{2}},...,{\bf
\Gamma}_{a_{1}a_{2} \cdot \cdot \cdot a_{D}}, \label{gamma-algebra}
\end{equation}
where ${\bf \Gamma }_{a_{1}a_{2}\cdot \cdot \cdot
a_{k}}=\frac{1}{k!}({\bf \Gamma}_{a_{1}}{\bf \Gamma }_{a_{2}}
\cdot \cdot \cdot {\bf \Gamma}_{a_{k}}\pm [$permutations$])$. For
even $D$ these are all linearly independent, but for odd $D$ they
are not, because ${\bf \Gamma }_{12\cdot \cdot \cdot D}=\sigma
{\bf I}$ and therefore half of them are proportional to the other
half. Thus, the dimension of this algebra is $m^{2}=2^{2[D/2]}$
and not $D^{2}$ as one could naively think. This representation
provides an elegant way to generate all $m\times m$ matrices (note
however, that $m=2^{[D/2]}$ is not any number).

\subsection{The Fermionic Generators}

The simplest extension of the matrices (\ref{ja}, \ref{jab}) is obtained by the
addition of one row and one column. The generators associated to these entries
would have one on spinor index. Let us call ${\bf Q}_{\gamma }$ the generator
that has only one nonvanishing entry in the $\gamma $-th row of the last
column,
\begin{equation}
{\bf Q}_{\gamma }=\left[
\begin{array}{cc}
0 & \delta _{\gamma }^{\alpha } \\
-C_{\gamma \beta } & 0
\end{array}
\right] .  \label{q-generator}
\end{equation}

Since this generator carries a spinorial index, we will assume it is in a spin
1/2 representation of the Lorentz group. The entries of the bottom row will be
chosen so as to produce smallest supersymmetric extensions of adS. There are
essentially two ways of reducing the representation compatible with Lorentz
invariance: {\em chirality}, which corresponds to Weyl spinors, and {\em
reality}, for {\em Majorana} spinors. A Majorana spinor satisfies a constraint
that relates its components to those of its complex conjugate,
\begin{equation}
\bar{\psi}^{\alpha }=C^{\alpha \beta }\psi _{\beta }.
\label{conjugation}
\end{equation}

The charge conjugation matrix, ${\bf C}=(C^{\alpha \beta })$ is invertible,
$C_{\alpha \beta}C^{\beta \gamma }=\delta _{\alpha }^{\gamma }$ and therefore,
it can be used as a metric in the space of Majorana spinors. Since both ${\bf
\Gamma }^{a}$ and $({\bf \Gamma}^{a})^{\rm{T}}$ obey the same Clifford algebra
($\{{\bf \Gamma }^{a},{\bf \Gamma }^{b}\}=2\eta ^{ab}$), there could be a
representation in which the $({\bf \Gamma }^{a})^{{\rm T}}$ is related to ${\bf
\Gamma}^{a}$ by a change of basis up to a sign,
\begin{equation}
({\bf \Gamma }^{a})^{\rm{T}}=\eta {\bf C\Gamma}^{a}{\bf C}^{-1}
\;\;\mbox{with}\ \eta ^{2}=1. \label{c-gamma}
\end{equation}

The Dirac matrices for which there is an operator ${\bf C}$ satisfying
(\ref{c-gamma}) is called the Majorana representation\footnote{Chirality is
defined only for even $D$, while the Majorana reality condition can be
satisfied in any $D$, provided the spacetime signature is such that, if there
are $s$ spacelike and $t$ timelike dimensions, then $s-t=0,1,2,6,7$ mod $8$
\cite{Sohnius,Freund} (that is $D=2,3,4,8,9,$ mod $8$ for lorentzian
signature). Thus, only in the latter case Majorana spinors can be defined
unambiguously.}. This last equation is the defining relation for the charge
conjugation matrix, and whenever it exists, it can be chosen to have definite
parity,
\begin{equation}
{\bf C}^{\rm{T}}=\lambda {\bf C,}\rm{ with }\lambda =\pm 1.
\label{c}
\end{equation}
It can be seen that with the choice (\ref{q-generator}), Majorana
conjugate of ${\bf \bar{Q}}$ is
\begin{eqnarray}
{\bf \bar{Q}}^{\gamma } &=&:C^{\alpha \beta }{\bf Q}_{\beta } \nonumber \\
&=&\left[
\begin{array}{cc}
0 & C^{\alpha \gamma } \\
-\delta _{\beta }^{\gamma } & 0
\end{array}
\right] .  \label{qbar-generator}
\end{eqnarray}

\subsection{Closing the Algebra}

We already encountered the bosonic generators responsible for the AdS
transformations (\ref{ja}, \ref{jab}), which has the general form required by
(\ref{bb}). It is also straightforward to check that commutators of the form
$[{\bf J},{\bf Q}]$ turn out to be proportional to ${\bf Q}$, in agreement with
the general form (\ref{bf}). What is by no means trivial is the closure of the
anticommutator $\{{\bf Q},{\bf Q}\}$ as in (\ref{ff}). Direct computation
yields

\begin{eqnarray}
\{{\bf Q}_{\gamma },{\bf Q}_{\lambda }\}_{\beta }^{\alpha } &=&\left[
\begin{array}{cc}
0 & \delta _{\gamma }^{\alpha } \\
-C_{\gamma \rho } & 0
\end{array}
\right] \left[
\begin{array}{cc}
0 & \delta _{\lambda }^{\rho } \\
-C_{\lambda \beta } & 0
\end{array}
\right] +(\gamma \leftrightarrow \lambda ) \\
&=&-\left[
\begin{array}{cc}
\delta _{\gamma }^{\alpha }C_{\lambda \beta }+\delta_{\lambda
}^{\alpha}C_{\gamma \beta } & 0 \\
0 & C_{\gamma \lambda}+C_{\lambda \gamma }
\end{array}
\right] .  \label{qq+qq}
\end{eqnarray}

The form of the lower diagonal piece immediately tells us that unless
$C_{\gamma \lambda}$ is antisymmetric, it will be necessary to include at least
one more bosonic generator (and possibly more) with nonzero entries in this
diagonal block. This relation also shows that the upper diagonal block is a
collection of matrices ${\bf M}_{\gamma \lambda }$ whose components are
\[
\left( M_{\gamma \lambda }\right)_{\beta }^{\alpha}=-(\delta_{\gamma}^{\alpha }
C_{\lambda \beta }+ \delta _{\lambda}^{\alpha}C_{\gamma \beta}).
\]
Multiplying both sides of this relation by $C$, one finds
\begin{equation}
\left( CM_{\gamma \lambda }\right) _{\alpha \beta }=-(C_{\alpha \gamma
}C_{\lambda \beta }+C_{\alpha \lambda }C_{\gamma \beta }),  \label{cc+cc}
\end{equation}
which is symmetric in $(\alpha \beta )$. This means that the bosonic generators
can only include those matrices in the Dirac algebra such that, when multiplied
by ${\bf C}$ on the left (${\bf CI},$ ${\bf C\Gamma }_{a},$ ${\bf C\Gamma
}_{a_{1}a_{2}},...,$ ${\bf C\Gamma }_{a_{1}a_{2}\cdot \cdot \cdot a_{D}}$) turn
out to be symmetric. The other consequence of this is that, if one wants to
have the AdS algebra as part of the superalgebra, both ${\bf C\Gamma}_{a}$ and
${\bf C\Gamma}_{ab}$ should be symmetric matrices. Now, multiplying
(\ref{c-gamma}) by ${\bf C}$ from the right, we have
\begin{equation}
({\bf C\Gamma }_{a})^{\rm{T}}=\lambda \eta {\bf C\Gamma }_{a},
\label{c-gamma2}
\end{equation}
which means that we need
\begin{equation}
\lambda \eta =1.  \label{eta}
\end{equation}
It can be seen that
\[
({\bf C\Gamma }_{ab})^{\rm{T}}=-\lambda {\bf C\Gamma }_{ab},
\]
which in turn requires
\[
\lambda =-1=\eta .
\]
This means that ${\bf C}$ is antisymmetric ($\lambda =-1$) and then the lower
diagonal block in (\ref{qq+qq}) vanishes identically. However, the values of
$\lambda $ and $\eta $ cannot be freely chosen but are fixed by the spacetime
dimension as is shown in the following table (see Ref.\cite{TrZ2} for details)
\[
\begin{array}{|c|c|c|}
\hline {\bf D} & \lambda & \eta \\ \hline
 3 & -1 & -1 \\
 5 & -1 & +1 \\
 7 & +1 & -1 \\
 9 & +1 & +1 \\
 11 &-1 & -1 \\ \hline
\end{array}
\]
and the pattern repeats mod $8.$ This table shows that the simple cases occur
for dimensions $3$ mod $8$, while for the remaining cases life is a little
harder. For $D=7$ mod $8$ the need to match the lower diagonal block with some
generators can be satisfied quite naturally by including several spinors
labeled with a new index, $\psi _{i}^{\alpha},i=1,...N$, and the generator of
supersymmetry should also carry the same index. This means that there are
actually $N$ supercharges or, as it is usually said, the theory has an extended
supersymmetry ($N\geq 2$). For $D=5$ mod $4$ instead, the superalgebra can be
made to close in spite of the fact that $\eta =+1$ if one allows complex spinor
representations, which is a particular form of extended supersymmetry since now
${\bf Q}_{\gamma}$ and ${\bf \bar{Q}}^{\gamma}$ are independent.

So far we have only given some restrictions necessary to close the algebra so
that the AdS generators appear in the anticommutator of two supercharges. In
general, however, apart from ${\bf J}_{a}$ and ${\bf J}_{ab}$ other matrices
will occur in the r.h.s. of the anticommutator of ${\bf Q}$ and ${\bf \bar{Q}}$
which extends the AdS algebra into a larger bosonic algebra. This happens even
in the cases where there is no extended supersymmetry ($N=1$). The bottom line
of this construction is that the supersymmetric extension of the AdS algebra
for each odd dimension falls into three different families:

$D=3$ mod $8$ (Majorana representation, $N\geq 1$),

$D=7$ mod $8$ (Majorana representation, even $N$), and

$D=5$ mod $4$ (complex representations, $N\geq 1$ [or $2N$ real spinors]).

The corresponding superalgebras\footnote{The algebra $osp(p|q)$ (resp.
$usp(p|q)$) is that which generates the orthosymplectic (resp.
unitary-symplectic) Lie group. This group is defined as the one that leaves
invariant the quadratic form $G_{AB}z^A z^B = g_{ab}x^a x^b + \gamma_{\alpha
\beta} \theta^{\alpha} \theta^{\beta}$, where $g_{ab}$ is a $p$-dimensional
symmetric (resp. hermitean) matrix and $\gamma_{\alpha \beta}$ is a
$q$-dimensional antisymmetric (resp. anti-hermitean) matrix.}were computed by
van Holten and Van Proeyen for $D=2, 3, 4$ mod $8$ in Ref. \cite{vH-VP}, and in
the other cases, in Refs.\cite{TrZ1,TrZ2}:

\begin{center}
\begin{tabular}{|l|c|c|}
\hline D & S-Algebra & Conjugation Matrix \\ \hline
3 mod $8$ & $osp(m|N)$ & $C^{T}=-C$ \\ \hline
7 mod $8$ & $osp(N|m)$ & $C^{T}=C$ \\ \hline
5 mod $4$ & $usp(m|N)$ & $C^{\dag }=C$ \\ \hline
\end{tabular}

\smallskip
\end{center}

\section{CS Supergravity Actions}

The supersymmetric extension of a given Lie algebra is a mathematical problem
that has a mathematical solution, as is known from the general studies of
superalgebras \cite{Kac}. A particularly interesting aspect of these algebras
is their representations. The previous discussion was devoted to that point, of
which some cases had been studied more than 20 years ago in Ref. \cite{vH-VP}.
What is not at all trivial is how to construct a field theory action that
reflects this symmetry.

We saw in the previous lecture how to construct CS actions for the AdS
connection for any $D=2n+1$. The question is now, how to repeat this
construction for the connection of a larger algebra in which AdS is embedded.
The solution to this problem is well known. Consider an arbitrary connection
one form ${\Bbb A}$, with values in some Lie algebra ${\mathfrak g}$, whose
curvature is ${\Bbb F}=d{\Bbb A}+{\Bbb A}\wedge {\Bbb A}$. Then, the $2n$-form
\begin{equation}
{\Bbb C}_{2n}\equiv <{\Bbb F}\wedge \cdot \cdot \cdot \wedge {\Bbb
F}>, \label{topclass}
\end{equation}
where $<\cdot \cdot \cdot >$ stands for an invariant trace, is invariant under
the group whose Lie algebra is ${\mathfrak g}$. Furthermore, ${\Bbb C}_{2n} $
is closed: $d{\Bbb C}_{2n}=0$, and therefore can be locally written as an exact
form,
\[
{\Bbb C}_{2n}=d{\Bbb L}_{2n-1}.
\]
The $(2n-1)$-form ${\Bbb L}_{2n-1}$ is a CS lagrangian, and therefore the
problem reduces to finding the invariant trace $<\cdot \cdot \cdot >$. The
canonical --and possibly unique-- choice of invariant trace with the features
required here is the {\bf supertrace}, which is defined as follows: if a matrix
has the form
\[
{\Bbb M}=\left[
\begin{array}{cc}
J_{b}^{a} & F_{\beta }^{a} \\
H_{b}^{\alpha } & S_{\beta }^{\alpha }
\end{array}
\right] ,
\]
where $a,b$ are (bosonic) tensor indices and $\alpha, \beta$ are (fermionic)
spinor indices, then $STr[{\Bbb M}]=Tr[{\bf J}]-Tr[{\bf S}
]=J_{a}^{a}-S_{\alpha}^{\alpha}$.

If we call ${\Bbb G}_{M}$ the generators of the Lie algebra, so that ${\Bbb
A}=G_{M}A^{M}$, ${\Bbb F}=G_{M}F^{M}$, then
\begin{eqnarray}
{\Bbb C}_{2n} &=&STr\left[{\Bbb G}_{M_{1}}\cdot \cdot \cdot {\Bbb
G}_{M_{n}}\right] F^{M_{1}}\cdot \cdot \cdot F^{M_{n}}  \nonumber
\\ &=&g_{_{\rm{M}_{1}\cdot \cdot \cdot \rm{M}_{n}}}F^{M_{1}}\cdot
\cdot \cdot F^{M}=d{\Bbb L}_{2n-1},  \label{defCS}
\end{eqnarray}
where $g_{_{\rm{M}_{1}\cdot \cdot \cdot \rm{M}_{n}}}$ is an invariant tensor of
rank $n$ in the Lie algebra. Thus, the steps to construct the CS lagrangian are
straightforward: Take the supertrace of all products of generators in the
superalgebra and solve equation (\ref{defCS}) for ${\Bbb L}_{2n-1}$. Since the
superalgebras are different in each dimension, the CS lagrangians differ in
field content and dynamical structure from one dimension to the next, although
the invariance properties are similar in all cases. The action
\begin{equation}
I_{2n-1}^{CS}[{\Bbb A}]=\int {\Bbb L}_{2n-1}  \label{generalCSaction}
\end{equation}
is invariant, up to surface terms, under the local gauge transformation
\begin{equation}
\delta {\Bbb A}=\nabla {\bf \Lambda},  \label{gaugetransf}
\end{equation}
where ${\bf \Lambda}$ is a zero-form with values in the Lie algebra ${\mathfrak
g}$, and $\nabla $ is the exterior covariant derivative in the representation
of ${\Bbb A}$. In particular, under a supersymmetry transformation, ${\bf
\Lambda}= \bar{\epsilon}^{i}Q_{i}-\bar{Q}^{i}\epsilon_{i}$, and
\begin{equation}
\delta _{\epsilon }{\Bbb A}=\left[
\begin{array}{cc}
\epsilon ^{k}\bar{\psi}_{k}-\psi ^{k}\bar{\epsilon}_{k} & D\epsilon_{j} \\
-D\bar{\epsilon}^{i} & \bar{\epsilon}^{i}\psi _{j}-\bar{\psi}^{i}\epsilon
_{j}
\end{array}
\right] ,  \label{delA}
\end{equation}
where $D$ is the covariant derivative on the bosonic connection,
\[
D\epsilon _{j}=\left( d+\frac{1}{2}[e^{a}{\bf \Gamma }_{a}+
\frac{1}{2}\omega ^{ab}{\bf \Gamma }_{ab}+\frac{1}{r!}b^{[r]}{\bf
\Gamma }_{[r]}]\right) \epsilon _{j}-a_{j}^{i}\epsilon _{i}.
\]
Two interesting cases can be mentioned here: \vskip 0.5cm
\begin{center}
A. {\bf D=5 SUGRA}
\end{center}

In this case the supergroup is $U(2,2|N)$. The associated connection can be
written as
\begin{equation}
{\Bbb A}{\bf =}e^{a}{\bf J}_{a}+\frac{1}{2}\omega^{ab} {\bf J}_{ab}+ A^{K} {\bf
T}_{K}+(\bar{\psi}^{r}{\bf Q}_{r}-{\bf \bar{Q}}^{r}\psi _{r})+A{\bf Z},
\label{D=5connection}
\end{equation}

where the generators ${\bf J}_{a}$, ${\bf J}_{ab}$, form an AdS algebra
($so(4,2)$), ${\bf T}_{K}$ ($K=1,\cdot \cdot \cdot N^{2}-1$) are the generators
of $su(N)$, ${\bf Z}$ generates a $U(1)$ subgroup and ${\bf Q}, {\bf \bar{Q}}$
are the supersymmetry generators, which transform in a vector representation of
$SU(N)$. The Chern-Simons Lagrangian for this gauge algebra is defined by the
relation $dL=iSTr[{\Bbb F}^{3}]$, where ${\Bbb F}=d{\Bbb A} + {\Bbb A}^{2}$ is
the (antihermitean) curvature. Using this definition, one obtains the
Lagrangian originally discussed by Chamseddine in \cite{Chamseddine},

\begin{equation}
L=L_{G}(\omega
^{ab},e^{a})+L_{su(N)}(A_{s}^{r})+L_{u(1)}(\omega^{ab},e^{a},A)+
L_{F}(\omega ^{ab},e^{a},A_{s}^{r},A,\psi _{r}),  \label{L}
\end{equation}
with
\begin{equation}
\begin{array}{lll}
L_{G} & = & \frac{1}{8}\epsilon_{abcde} \left[ R^{ab}R^{cd} e^{e}/l +
\frac{2}{3}R^{ab}e^{c}e^{d}e^{e}/l^{3} + \frac{1}{5}
e^{a}e^{b}e^{c}e^{d}e^{e}/l^{5} \right] \\
L_{su(N)} & = &-Tr\left[ {\bf A}(d{\bf A})^{2}+\frac{3}{2}{\bf A}^{3}d{\bf A }+
\frac{3}{5}{\bf A}^{5}\right] \\
L_{u(1)} & = & \left( \frac{1}{4^{2}}-\frac{1}{N^{2}}\right)
A(dA)^{2}+\frac{ 3}{4l^{2}}\left[
T^{a}T_{a}-R^{ab}e_{a}e_{b}-l^{2}R^{ab}R_{ab}/2\right] A \\
&  & +\frac{3}{N}F_{s}^{r}F_{r}^{s}A \\
L_{f} & = &
\begin{array}{l}
\frac{3}{2i}\left[ \bar{\psi}^{r}{\cal R}\nabla \psi
_{r}+\bar{\psi}^{s} {\cal F}_{s}^{r}\nabla \psi _{r}\right] +c.c.
\end{array}
\end{array}
,  \label{Li}
\end{equation}
where $A_{s}^{r}\equiv A^{K}({\bf T}_{K})_{s}^{r}$ is the $su(N)$ connection,
$F_{s}^{r}$ is its curvature, and the bosonic blocks of the supercurvature:
${\cal R}=\frac{1}{2}T^{a}{\bf \Gamma}_{a} + \frac{1}{4}
(R^{ab}+e^{a}e^{b}){\bf \Gamma}_{ab}+ \frac{i}{4}dA{\bf I}-\frac{1}{2}\psi
_{s}\bar{\psi}^{s}$, ${\cal F}_{s}^{r}=F_{s}^{r}+\frac{i}{N}dA\delta_{s}^{r} -
\frac{1}{2}\bar{\psi}^{r}\psi _{s}$. The cosmological constant is $ -l^{-2},$
and the AdS covariant derivative $\nabla $ acting on $\psi _{r}$ is
\begin{equation}
\nabla \psi _{r}=D\psi _{r}+\frac{1}{2l}e_{a}^{a}{\bf \Gamma}\psi
_{r}-A_{\,r}^{s}\psi _{s}+i\left( \frac{1}{4}-\frac{1}{N}\right) A\psi _{r}.
\label{delta}
\end{equation}
where $D$ is the covariant derivative in the Lorentz connection.

The above relation implies that the fermions carry a $u(1)$ ``electric'' charge
given by $e=\left( \frac{1}{4}- \frac{1}{N}\right)$. The purely gravitational
part, $L_{G}$ is equal to the standard Einstein-Hilbert action with
cosmological constant, plus the dimensionally continued Euler
density\footnote{The first term in $L_{G}$ is the dimensional continuation of
the Euler (or Gauss-Bonnet) density from two and four dimensions, exactly as
the three-dimensional Einstein-Hilbert Lagrangian is the continuation of the
the two dimensional Euler density. This is the leading term in the limit of
vanishing cosmological constant ($l\rightarrow \infty )$, whose local
supersymmetric extension yields a nontrivial extension of the Poincar\'{e}
group \cite{BTrZ}.}.

The action is by construction invariant --up to a surface term-- under the
local (gauge generated) supersymmetry transformations $\delta_{\Lambda}{\Bbb
A}=-(d{\bf \Lambda }+ [{\Bbb A},{\bf \Lambda}])$ with ${\bf \Lambda
}=\bar{\epsilon}^{r}{\bf Q}_{r}-{\bf \bar{Q}}^{r}\epsilon _{r}$, or

\[
\begin{array}{lll}
\delta e^{a} & = & \frac{1}{2}\left( \overline{\epsilon }^{r}{\bf \Gamma }
^{a}\psi _{r}-\bar{\psi}^{r}{\bf \Gamma }^{a}\epsilon _{r}\right) \\
\delta \omega ^{ab} & = & -\frac{1}{4}\left( \bar{\epsilon}^{r}{\bf \Gamma }
^{ab}\psi _{r}-\bar{\psi}^{r}{\bf \Gamma }^{ab}\epsilon _{r}\right) \\
\delta A_{\,s}^{r} & = & -i\left( \bar{\epsilon}^{r}\psi_{s}-\bar{\psi}
^{r}\epsilon _{s}\right) \\
\delta \psi _{r} & = & -\nabla \epsilon _{r} \\
\delta \bar{\psi}^{r} & = & -\nabla \bar{\epsilon}^{r} \\
\delta A & = & -i\left( \bar{\epsilon}^{r}\psi _{r}-\bar{\psi}^{r}\epsilon
_{r}\right) .
\end{array}
\]

As can be seen from (\ref{Li}) and (\ref{delta}), for $N=4$ the $U(1)$ field
$A$ looses its kinetic term and decouples from the fermions (the gravitino
becomes uncharged with respect to $U(1)$). The only remnant of the interaction
with the $A$ field is a dilaton-like coupling with the Pontryagin four forms
for the AdS and $SU(N)$ groups (in the bosonic sector). As it is shown in
Ref.\cite{ChTrZ}, the case $N=4$ is also special at the level of the algebra,
which becomes the superalgebra $su(2,2|4)$ with a $u(1)$ central extension.

In the bosonic sector, for $N=4$, the field equation obtained from the
variation with respect to $A$ states that the Pontryagin four form of AdS and
$SU(N)$ groups are proportional. Consequently, if the spatial section has no
boundary, the corresponding Chern numbers must be related. Since $\Pi
_{4}(SU(4))=0$, the above implies that the Pontryagin plus the Nieh-Yan number
must add up to zero.

\vskip 0.5cm
\begin{center}
B. {\bf D=11 SUGRA}
\end{center}

In this case, the smallest AdS superalgebra is $osp(32|1)$ and the connection
is
\begin{equation}
{\Bbb A}=\frac{1}{2}\omega ^{ab}J_{ab}+e^{a}J_{a}+\frac{1}{5!} A^{abcde}
J_{abcde}+\bar{Q}\psi , \label{D=11 connection}
\end{equation}
where $A^{abcde}$ is a totally antisymmetric fifth-rank Lorentz tensor
one-form. Now, in terms of the elementary bosonic and fermionic fields, the CS
form in ${\Bbb L}_{2n-1}$reads
\begin{equation}
{\Bbb L}_{11}^{osp(32|1)}({\Bbb A})=L_{11}^{sp(32)}({\bf \Omega})+L_{f}({\bf
\Omega },\psi ), \label{L11}
\end{equation}
where ${\bf \Omega }\equiv \frac{1}{2}(e^{a}{\bf \Gamma}_{a}+\frac{1}{2}
\omega^{ab}{\bf \Gamma}_{ab}+ \frac{1}{5!}A^{abcde}{\bf \Gamma}_{abcde})$ is an
$sp(32)$ connection. The bosonic part of (\ref{L11}) can be written as
\[
L_{11}^{sp(32)}({\bf \Omega
})=2^{-6}L_{G\;11}^{AdS}(\omega,e)-\frac{1}{2}
L_{T\;11}^{AdS}(\omega ,e)+L_{11}^{b}(A,\omega ,e),
\]
where $L_{G\;11}^{AdS}$ is the CS form associated to the $12$-dimensional Euler
density, and $L_{T\;11}^{AdS}$ is the CS form whose exterior derivative is the
Pontryagin form for $SO(10,2)$ in $12$ dimensions. The fermionic Lagrangian is
\begin{eqnarray*}
L_{f} &=&6(\bar{\psi}{\cal R}^{4}D\psi )-3\left[
(D\bar{\psi}D\psi) + (\bar{\psi}{\cal R}\psi
)\right] (\bar{\psi}{\cal R}^{2}D\psi ) \\
&&-3\left[ (\bar{\psi}{\cal R}^{3}\psi
)+(D\bar{\psi}{\cal R}^{2}D\psi )\right] (\bar{\psi}D\psi )+ \\
&&2\left[ (D\bar{\psi}D\psi )^{2}+(\bar{\psi}{\cal R}\psi
)^{2}+(\bar{\psi} {\cal R}\psi)(D\bar{\psi}D\psi )\right]
(\bar{\psi}D\psi ),
\end{eqnarray*}
where ${\cal R}=d{\bf \Omega }+{\bf \Omega }^{2}$ is the $sp(32)$
curvature. The supersymmetry transformations (\ref{delA}) read
\[
\begin{array}{lll}
\delta e^{a}=\frac{1}{8}\bar{\epsilon}{\bf \Gamma }^{a}\psi & \hspace{1cm} &
\delta \omega ^{ab}=-\frac{1}{8}\bar{\epsilon}{\bf \Gamma }^{ab}\psi \\ &  &
\\ \delta \psi =D\epsilon & \hspace{1cm} & \delta A^{abcde}=\frac{1}{8}\bar{
\epsilon}{\bf \Gamma }^{abcde}\psi .
\end{array}
\]

Standard (CJS) eleven-dimensional supergravity \cite{CJS} is an N=1
supersymmetric extension of Einstein-Hilbert gravity that cannot admit a
cosmological constant \cite{BDHS,Deser}. An $N>1$ extension of the CJS theory
is not known. In our case, the cosmological constant is necessarily nonzero by
construction and the extension simply requires including an internal $so(N)$
gauge field coupled to the fermions. The resulting Lagrangian is an $osp(32|N)$
CS form \cite{Troncoso}.

\section{Summary}

The supergravities presented here have two distinctive features: The
fundamental field is always the connection ${\Bbb A}$ and, in their simplest
form, they are pure CS systems (matter couplings are discussed below). As a
result, these theories possess a larger gravitational sector, including
propagating spin connection. Contrary to what one could expect, the geometrical
interpretation is quite clear, the field structure is simple and, in contrast
with the standard cases, the supersymmetry transformations close off shell
without auxiliary fields.

{\bf Torsion.} It can be observed that the torsion Lagrangians, $L_{T}$, are
odd while the torsion-free terms, $L_{G}$, are even under spacetime
reflections. The minimal supersymmetric extension of the AdS group in $4k-1$
dimensions requires using chiral spinors of $SO(4k)$ \cite{Gunaydin}. This in
turn implies that the gravitational action has no definite parity and requires
the combination of $L_{T}$ and $L_{G}$ as described above. In $D=4k+1$ this
issue doesn't arise due to the vanishing of the torsion invariants, allowing
constructing a supergravity theory based on $L_{G}$ only, as in
\cite{Chamseddine}. If one tries to exclude torsion terms in $4k-1$ dimensions,
one is forced to allow both chiralities for $SO(4k)$ duplicating the field
content, and the resulting theory has two copies of the same system
\cite{Horava}.

{\bf Field content and extensions with N$>$1.} The field content compares with
that of the standard supergravities in $D=5,7,11$ in the following table, which
shows the corresponding supergravities

\begin{center}
$
\begin{array}{|c|c|c|c|}
\hline D & {\rm Standard ~supergravity} & {\rm CS ~supergravity} &
{\rm Algebra}\\ \hline 5 & e_{\mu}^{a}\;\psi _{\mu }^{\alpha
}\;\bar{\psi}_{\alpha \mu } & e_{\mu }^{a}\;\omega _{\mu
}^{ab}\;A_{\mu }\;A_{j\mu }^{i}\;\psi _{i\mu }^{\alpha
}\;\bar{\psi}_{\alpha \mu }^{i}, \;i,j=1,...N & usp(2,2|N)\\
\hline 7 & e_{\mu }^{a}\;A_{[3]}\;a_{\mu j}^{i}\;\lambda ^{\alpha
}\;\phi \;\psi _{\mu }^{\alpha i} & e_{\mu }^{a}\;\omega _{\mu
}^{ab}\;A_{\mu j}^{i}\;\psi _{\mu }^{\alpha i},\; i,j=1,...N=2n &
osp(N|8) \\  \hline 11 & e_{\mu }^{a}\;A_{[3]}\;\psi _{\mu
}^{\alpha } & e_{\mu }^{a} \;\omega _{\mu }^{ab}\;A_{\mu
}^{abcde}\;\psi _{\mu }^{\alpha }\;,\;i,j=1,...N\; & osp(32|N) \\
\hline
\end{array}
$
\end{center}

Standard supergravity in five dimensions is dramatically different from the
theory presented here, which was also discussed by Chamseddine in
\cite{Chamseddine}.

Standard seven-dimensional supergravity is an $N=2$ theory (its maximal
extension is $N=4$), whose gravitational sector is given by Einstein-Hilbert
gravity with cosmological constant and with a background invariant under
$OSp(2|8)$ \cite{D=7,Salam-Sezgin}. Standard eleven-dimensional supergravity
\cite{CJS} is an $N=1$ supersymmetric extension of Einstein-Hilbert gravity
with vanishing cosmological constant. An $N>1$ extension of this theory is not
known.

In our construction, the extensions to larger $N$ are straightforward in any
dimension. In $D=7$, the index $i$ is allowed to run from $2$ to $2s$, and the
Lagrangian is a CS form for $osp(2s|8)$. In $D=11$, one must include an
internal $so(N)$ field and the Lagrangian is an $osp(32|N)$ CS form
\cite{TrZ1,TrZ2}. The cosmological constant is necessarily nonzero in all
cases.

{\bf Spectrum.} The stability and positivity of the energy for the solutions of
these theories is a highly nontrivial problem. As shown in Ref. \cite{BGH}, the
number of degrees of freedom of bosonic CS systems for $D\geq 5$ is not
constant throughout phase space and different regions can have radically
different dynamical content. However, in a region where the rank of the
symplectic form is maximal the theory may behave as a normal gauge system, and
this condition would be stable under perturbations. As it is shown in
\cite{ChTrZ} for $D=5$, there exists a nontrivial extension of the AdS
superalgebra with a central extension in anti-de Sitter space with only a
nontrivial $U(1)$ connection but no other matter fields. In this background the
symplectic form has maximal rank and the gauge superalgebra is realized in the
Dirac brackets. This fact ensures a lower bound for the mass as a function of
the other bosonic charges \cite{GH}.

{\bf Classical solutions.} The field equations for these theories, in terms of
the Lorentz components ($\omega $, $e$, $A$, ${\bf A}$, $\psi $), are the
different Lorentz tensor components for $<{\Bbb F}^{n-1}{\Bbb G}_{M}>=0$. It is
rather easy to verify that in all these theories the anti-de Sitter space is a
classical solution , and that for $\psi =A={\bf A}=0$ there exist spherically
symmetric, asymptotically AdS standard \cite{JJG}, as well as topological black
holes \cite{ABHPB}. In the extreme case these black holes can be shown to be
BPS states \cite{AMTrZ}.

{\bf Matter couplings.} It is possible to introduce minimal couplings to matter
of the form ${\Bbb A}\cdot {\Bbb J}^{ext}$. For $D=5$, the theory couples to an
electrically charged $U(1)$ 0 brane (point charge), to $SU(4)$ -colored 0
branes (quarks) or to uncharged 2-brane, whose respective worldhistories couple
to $A_{\mu }$, ${\bf A}_{\mu }^{rs}$ and $\omega _{\mu }^{ab}$ respectively.
For $D=11$, the theory admits a 5-brane and a 2-brane minimally coupled to
$A_{\mu }^{abcde}$ and \thinspace $\omega _{\mu }^{ab}$ respectively.

{\bf Standard SUGRA.} Some sector of these theories might be related to the
standard supergravities if one identifies the totally antisymmetric part of
$\omega_{\mu }^{ab}$ in a coordinate basis, $k_{\mu \nu \lambda }$, (sometimes
called the contorsion tensor) with the abelian 3-form, $A_{[3]}$. In 11
dimensions one could also identify the totally antisymmetrized part of $A_{\mu
}^{abcde}$ with an abelian 6-form $A_{[6]}$, whose exterior derivative,
$dA_{[6]}$, is the dual of $F_{[4]}=dA_{[3]}$. Hence, in $D=11$ the CS theory
may contain the standard supergravity as well as some kind of dual version of
it.

{\bf Gravity sector.} A most remarkable result from imposing the supersymmetric
extension, is the fact that if one sets all fields, except those that describe
the geometry --$e^{a}$ and $\omega ^{ab}$-- to zero, the remaining action has
no free parameters. This means that the gravity sector is uniquely fixed. This
is remarkable because as we saw already for $D=3$ and $D=7$, there are several
CS actions that one can construct for the AdS gauge group, the Euler CS form
and the so-called exotic ones, that include torsion explicitly, and the
coefficients for these different CS lagrangians is not determined by the
symmetry considerations. So, even from a purely gravitational point of view, if
the theory admits a supersymmetric extension, it has more predictive power than
if it does not.

\vskip 0.7 cm
\begin{center}
{\bf LECTURE 4} \vskip 0.3 cm {\bf EPILOGUE: DYNAMICAL CONTENT of
CHERN SIMONS THEORIES}
\end{center}
\vskip 0.5 cm

The physical meaning of a theory is defined by the dynamics it
displays both at the classical and quantum levels. In order to
understand the dynamical contents of the classical theory, the
physical degrees of freedom must be identified. In particular, it
should be possible --at least in principle-- to separate the
propagating modes from the gauge degrees of freedom, and from
those which do not evolve independently at all (second class
constraints). The standard way to do this is Dirac's constrained
Hamiltonian analysis and has been applied to CS systems in
\cite{BGH}. Here we summarize this analysis and refer the reader
to the original papers for details. It is however, fair to say
that a number of open problems remain and it is a area of research
which is at a very different stage of development compared with
the previous discussion.

\section{Hamiltonian Analysis}

From the dynamical point of view, a CS system can be described by
a Lagrangian of the form\footnote{ Note that in this section, for
notational simplicity, we assume the spacetime to be
($2n+1$)-dimensional.}
\begin{equation}
L_{2n+1}=l_{a}^{i}(A_{j}^{b})\dot{A}_{i}^{a}-A_{o}^{a}K_{a},
\end{equation}
where the ($2n+1$)-dimensional spacetime has been split into space and time, and
\[
K_{a}=-\frac{1}{2^{n}n}\gamma _{aa_{1}....a_{n}}\epsilon
^{i_{1}...i_{2n}}F_{i_{1}i_{2}}^{a_{1}}\cdot \cdot \cdot
F_{i_{2n-1}i_{2n}}^{a_{n}}.
\]
The field equations are
\begin{eqnarray}
\Omega _{ab}^{ij}(\dot{A}_{j}^{b}-D_{j}A_{0}^{b}) &=&0,  \label{csEq} \\
K_{a} &=&0,  \label{K=0}
\end{eqnarray}
where
\begin{eqnarray}
\Omega _{ab}^{ij} &=&\frac{\delta l_{b}^{j}}{\delta A_{i}^{a}}-\frac{\delta
l_{a}^{i}}{\delta A_{j}^{b}}  \label{symplectic} \\
&=&-\frac{1}{2^{n-1}}\gamma _{aba_{2}....a_{n}}\epsilon
^{iji_{3}...i_{2n}}F_{i_{3}i_{4}}^{a_{2}}\cdot \cdot \cdot
F_{i_{2n-1}i_{2n}}^{a_{n}}  \nonumber
\end{eqnarray}
is the {\bf symplectic form}. The passage to the Hamiltonian has
the problem that the velocities appear linearly in the Lagrangian
and therefore there are a number of primary constraints
\begin{equation}
\phi _{a}^{i}\equiv p_{a}^{i}-l_{a}^{i}\approx 0.  \label{Phi's}
\end{equation}

Besides these, there are secondary constraints $K_{a}\approx 0$,
which can be combined with the $\phi$s into the expressions
\begin{equation}
G_{a}\equiv -K_{a}+D_{i}\phi _{a}^{i}.  \label{Ga}
\end{equation}
The complete set of constraints forms a closed Poison bracket algebra,
\[
\begin{array}{ll}
\{\phi _{a}^{i},\phi _{b}^{j}\} & =\Omega _{ab}^{ij} \\
\{\phi _{a}^{i},G_{b}\} & =f_{ab}^{c}\phi _{c}^{i} \\
\{G_{a},G_{b}\} & =f_{ab}^{c}G_{c}
\end{array},
\]
where $f_{ab}^{c}$ are the structure constants of the gauge algebra of the
theory. Clearly the $G$s form a first class algebra which reflects the gauge
invariance of the theory, while some of the $\phi $s are second class and some
are first class, depending on the rank of the symplectic form $\Omega$.

\subsection{Degeneracy}
An intriguing aspect of Chern-Simons theories is the multiplicity of ground
states that they can have. This can be seen from the field equations, which for
$D=2n+1$, are polynomials of degree $n$ which in general have a very rich root
structure. As the symplectic form is field-dependent, the rank of the matrix
$\Omega _{ab}^{ij}$ need not be constant. It can change from one region of
phase space to another, with different degrees of degeneracy. Regions in phase
space with different degrees of degeneracy define dynamically distinct and
independent effective theories \cite{STZ}. If the system reaches a degenerate
configuration, some degrees of freedom are frozen in an irreversible process
which erases all traces of the initial conditions of the lost degrees of
freedom. One can speculate about the potential of this phenomenon as a way to
produce dimensional reduction through a dynamical process.

This issue was analyzed in the context of some simplified mechanical models and
the conclusion was that the degeneracy of the system occurs at submanifolds of
lower dimensionality in phase space, which are sets of unstable initial states
or sets of stable end points for the evolution \cite{STZ}. Unless the system is
chaotic, it can be expected that {\bf generic configurations}, where the rank
of $\Omega _{ab}^{ij}$ is maximal, fill most of phase space. As it was shown in
Ref. \cite{STZ}, if the system evolves along an orbit that reaches a surface of
degeneracy, $\Sigma $, it becomes trapped by the surfece and loses the degrees
o freedom that correspond to displacements away from $\Sigma $. This is an
irreversible process which can be viewed as mechanism for dynamical reduction
of degrees of freedom or dimensional reduction. A process of this type is seen
to take place in the dynamics of vortices in a fluid, where two vortices
coalesce and annihilate each other in an irreversible process.

\subsection{Generic counting}
There is a second problem and that is how to separate the first and second
class constraints among the $\phi $s. In Ref.\cite{BGH} the following results
are shown:
\begin{itemize}
\item The maximal rank of $\Omega _{ab}^{ij}$ is $2nN-2n$ , where $N$ is the
number of generators in the gauge Lie algebra.

\item There are $2n$ first class constraints among the $\phi$s which correspond
to the generators of spatial diffeomorphisms (${\cal H}_{i}$).

\item  The generator of timelike reparametrizations ${\cal H}_{\perp }$ is not
an independent first class constraint.
\end{itemize}

Putting all these facts together one concludes that, in a generic
configuration, the number of degrees of freedom of the theory ($\zeta ^{CS}$)
is
\begin{eqnarray}
\zeta ^{CS} &=&({\rm number} \:{\rm of} \: {\rm coordinates}) - ({\rm number}\:
{\rm of}\: {\rm 1st}\: {\rm class}\: {\rm constraints}) \nonumber \\
 & &-\frac{1}{2}({\rm number}\: {\rm  of}\: {\rm  2nd}\: {\rm class}\: {\rm constraints}) \nonumber \\
 &=&2nN-(N+2n)-\frac{1}{2}(2nN-2n) \label{nN-N-n} \\
 &=&nN-N-n.
\end{eqnarray}

This result is somewhat perplexing. A standard (metric) Lovelock theory of
gravity in $D=2n+1$ dimensions, has
\begin{eqnarray*}
\zeta ^{Lovelock} &=&D(D-3)/2 \\
&=&(2n+1)(n-1)
\end{eqnarray*}
propagating degrees of freedom \cite{Te-Z}. A CS gravity system for the AdS
group in the same dimension gives a much larger number,
\begin{equation}
\zeta ^{CS}=2n^{3}+n^{2}-3n-1.  \label{n3}
\end{equation}

In particular, for $D=5$, $\zeta ^{CS}=13$, while $\zeta ^{Lovelock}=5$. The
extra degrees of freedom correspond to propagating modes in $\omega ^{ab}$,
which in the CS theory are independent from the metric ones contained in
$e^{a}$.

As it is also shown in \cite{BGH}, an important simplification occurs when the
group has an invariant abelian factor. In that case the symplectic matrix
$\Omega _{ab}^{ij}$ takes a partially block-diagonal form where the kernel has
the maximal size allowed by a generic configuration. It is a nice surprise in
the cases of CS supergravities discussed above that for certain unique choices
of $N$, the algebras develop an abelian subalgebra and make the separation of
first and second class constraints possible (e.g., $N=4$ for $D=5$, and $N=32$
for $d=11$). In some cases the algebra is not a direct sum but an algebra with
an abelian central extension ($D=5$). In other cases, the algebra is a direct
sum, but the abelian subgroup is not put in by hand but it is a subset of the
generators that decouple from the rest of the algebra ($D=11$).

\subsection{Regularity conditions}
The counting discussed in \cite{BGH} was found to fail in the particular
example of CS supergravity in 5 dimensions. This is due to a different kind of
difficulty: the fact that the symmetry generators (first class constraints) can
fail to be functionally independent at some points of phase space. This is a
second type of degeneracy and makes it impossible to approximate the theory by
a linearized one. In fact, it can be seen that the number of degrees of freedom
of the linearized theory is larger than in the original one \cite{ChTrZ}. This
is the subject of an ongoing investigation which will be reported elsewhere
\cite{MiZ}.

\section{Final Comments}

{\bf 1.}Everything we know about the gravitational interaction at the classical
level, is described by Einstein's theory in four dimensions, which in turn is
supported by a handful of experimental observations. There are many
indications, however, that make it plausible to accept that our spacetime has
more dimensions than those that meet the eye. In a spacetime of more than four
dimensions, it is not logically necessary to consider the Einstein-Hilbert
action as the best description for gravity. In fact, string theory suggests a
Lanczos-Lovelock type action as more natural \cite{Zwiebach}. The large number
of free parameters in the LL action, however, cannot be fixed by arguments from
string theory. As we have shown, the only case in which there is a simple
symmetry principle to fix these coefficients is odd dimensions and that leads
to the Chern-Simons theories.

{\bf 2.}The CS theories of gravity have a profound geometrical meaning that
relates them to topological invariants --the Euler and the Chern or Pontryagin
classes-- and come about in a very natural way in a framework where the affine
and metric structures of the geometry are taken to be independent dynamical
objects. If one demands furthermore the theory to admit supersymmetry, there
is, in each dimensions essentially a unique extension which completely fixes
the gravitational sector, including the precise role of torsion in the action.

{\bf 3.}The CS theories of gravity obtained are classically and
semiclassically interesting. They possess nontrivial black hole solutions
\cite{BTZ94} which asymptotically approach spacetimes of constant negative
curvature (AdS spacetimes). These solutions have a thermodynamical behavior
which is unique among all possible black holes in competing LL theories with
the same asymptotics \cite{BHscan}. These black holes have positive and can
therefore always reach thermal equilibrium with their surroundings. These
theories also admit solutions which represent black objects, in the sense that
they possess a horizon that hides a singularity, but the horizon topology is
not spherical but a surface of constant nonpositive Ricci curvature
\cite{topolBH}. Furthermore, these solutions seem to have a well defined,
quantum mechanically stable ground states \cite{AMTrZ} which have been shown to
be BPS states of diverse topologies.

{\bf 4.}We have no way of telling at present what will be the fate of string
theory as a description of all interactions and constituents of nature. If it
is the right scenario and gravity is just a low energy effective theory that
would be a compelling reason to study gravity in higher dimensions, not as an
academic exercise as could have seemed in the time of Lanczos, but as a tool to
study big bang cosmology or black hole physics for instance. The truth is that
a field theory can tell us a lot a bout the low energy phenomenology, in the
same way that ordinary quantum mechanics tells us a lot about atomic physics
even if we know that is all somehow contained in QED.

{\bf 5.}Chern-Simons theories contain a wealth of other interesting features,
starting with their relation to geometry, gauge theories and knot invariants.
The higher-dimensional CS systems remain somewhat mysterious especially because
of the difficulties to treat them as quantum theories. However, they have many
ingredients that make CS theories likely models to be quantized: They carry no
dimensionful couplings, the only parameters they have are quantized, they are
the only ones in the Lovelock family of gravity theories that give rise to
black holes with positive specific heat \cite {BHscan} and hence, capable of
reaching thermal equilibrium with an external heat bath. Efforts to quantize CS
systems seem promising at least in the cases in which the space admits a
complex structure so that the symplectic form can be cast as a K\"{a}hler form
\cite{NS}. However, there is a number of open questions that one needs to
address before CS theories can be applied to describe the microscopic world,
like their Yang-Mills relatives. Until then, they are beautiful mathematical
models and interesting physical systems worth studying.

{\bf 6.}If the string scenario fails to deliver its promise, more work will
still be needed to understand the field theories it is supposed to represent,
in order to decipher their deeper interrelations. In this case, geometry is
likely to be an important clue, very much in the same way that it is an
essential element in Yang Mills and Einstein's theory. One can see the
construction discussed in these lectures as a walking tour in this direction.

It is perhaps appropriate to end these lectures quoting E. Wigner in full
\cite{Wigner}:

"The miracle of appropriateness of the language of mathematics for
the formulation of the laws of physics is a wonderful gift which
we neither understand nor deserve.  We should be grateful for it,
and hope that it will remain valid for future research, and that
it will extend, for better or for worse, to our pleasure even
though perhaps also to our bafflement, to wide branches of
learning''.

\vskip 1cm

\acknowledgments

It is a pleasure for me to thank the organizers and the staff of the school for
the stimulating discussions and friendly atmosphere in the charming colonial
city of Villa de Leyva. I am especially thankful for the efforts of M.
Kovacsics to make sure that a vegetarian would not only survive but be treated
with best fruits of the land. I wish to thank R. Aros, J. Bellissard, C.
Mart\'{\i}nez, S. Theisen, C. Teitelboim and R. Troncoso for many enlightening
discussions and helpful comments, and F. Mansouri for pointing out to me a
number of interesting references. I am especially grateful to Sylvie Paycha for
carefully proof reading and correcting many mistakes in the original
manuscript, both in style and in mathematics. I wish also thank the Albert
Einstein Institute for Gravitational Physics of the Max Planck Institute at
Potsdam for the hospitality, encouragement and resources to write up these
notes. This work was supported in part by FONDECYT grants 1990189, 1010450,
1020629 and by the generous institutional support to CECS from Empresas CMPC.
CECS is a Millennium Science Institute.

\end{document}